\documentclass[aps,pre,twocolumn,amsmath,amssymb,floatfix, showpacs]{revtex4-1}
\usepackage{graphicx}
\usepackage{dcolumn} 
\usepackage{bm}      
\usepackage[usenames,dvipsnames]{color}
\usepackage{ulem} 

\begin{document}
\title{Helicity dynamics in stratified turbulence in the absence of forcing}
\author{C. Rorai$^{1,2}$, D. Rosenberg$^1$,  A. Pouquet$^1$ and P.D. Mininni$^{1,3}$}
\affiliation{
 $^1$Computational and Information Systems Laboratory, NCAR, 
         P.O. Box 3000, Boulder CO 80307, USA. \\
 $^2$University of Trieste, Doctorate School in Environmental and Industrial
         Fluid Mechanics, Via Valerio, 12//b--34127 Trieste, Italy  \\
 $^3$Departamento de F\'\i sica, Facultad de Ciencias Exactas y
         Naturales,  \& IFIBA, CONICET, Ciudad 
         Universitaria, 1428 Buenos Aires, Argentina
         }

\begin{abstract}
A numerical study of decaying stably-stratified flows is performed. Relatively high stratification (Froude number $\sim 10^{-2}-10^{-1}$), and moderate Reynolds ($Re$) numbers ($Re \sim 3-6 \cdot 10^{3}$) are considered, and a particular emphasis is placed on the role of helicity (velocity-vorticity correlations), which is not an invariant of the non-dissipative equations. The problem is tackled by integrating the Boussinesq equations in a periodic cubical domain
using different initial conditions: a non-helical Taylor-Green (TG) flow, a fully
helical Beltrami (ABC) flow, and random flows with a tunable helicity. We show that for stratified ABC flows helicity undergoes a substantially slower decay than for  unstratified ABC flows. This fact is likely associated to the combined effect of stratification and large scale coherent structures. Indeed, when the latter are missing, as in random flows, helicity is rapidly destroyed by the onset of gravitational waves. A type of large-scale dissipative ``cyclostrophic'' balance
can be invoked to explain this behavior. No production of helicity is observed, contrary to the case of rotating and stratified flows. When helicity survives in the system it strongly affects the temporal energy decay and the energy distribution among Fourier modes. We discover in fact that the decay rate of energy for stratified helical flows is much slower than for stratified non-helical flows and can be considered with a phenomenological model in a way similar to what is done for unstratified rotating flows. We also show that helicity, when strong, has a measurable effect on the Fourier spectra, in particular at scales larger than the buoyancy scale  for which it displays a rather flat scaling associated with vertical shear.
\end{abstract}

\maketitle

\section{Introduction} \label{sec:intro}

Stratified flows, characterized by the  Brunt-Va\"iss\"al\"a frequency, $N$, are encountered in many astrophysical and geophysical settings, with the ratio of the gravity wave period $\tau_W=1/N$ to the eddy turn-over time $\tau_{NL}=L_0/U_0$ being rather small.
This ratio is the Froude number,  $Fr=U_0/[NL_0]$, where $U_0$ and $ L_0$ are the characteristic velocity and length scales of the fluid. For example, in the atmosphere, Froude numbers  of the order of $10^{-1}$ or $10^{-2}$ are encountered, whereas in the oceans they can be ten times smaller. Such flows, due to their deep connection to  
and their ubiquity in our environment, have been studied and reviewed extensively for their scaling and statistical properties  \cite{riley_rev_00, staquet_rev_02} and for their physical structures \cite{cambon_05}, as well as in the context of wind energy.
 
With $Fr<<1$, the waves are fast compared to turbulent eddies; they can lead to the formation of stratified layers with strong vertical gradients, to intensified mixing \cite{peltier_03}, in particular in the ocean  \cite{ivey_08}, and they can dominate the dynamics, at least at  scales larger than the so-called buoyancy scale $L_b$ at which $Fr=1$, namely $L_b=U_0/N$. This has led to the development of a suite of approximations, simpler and easier to integrate than the primitive equations, using either asymptotic expansions (see, e.g., \cite{embid_majda_98,  riley_rev_00, julien_06, majda_08, klein_rev_10, grooms_10, wingate_11, julien_12}), or  two-point closures of turbulence \cite{cambon_94, staquet_98, godef_03}, as well as weak turbulence statistical approaches \cite{nazar, newell, caillol}. 

As one considers the Froude number at scale $\ell$ with $u_{\ell}$ the velocity at that scale, the Brunt-Va\"iss\"al\"a frequency can be considered as  constant {\it vis \`a vis} scale dependency, whereas the eddy turn-over time, $\ell/u_{\ell}$, decreases with scale. As one moves to smaller scales, non-linear interactions are more effective and one can define the scale $\ell$ at which  $\tau_{NL}(\ell)=\tau_W(\ell)$; this scale is called the Ozmidov scale $L_{oz}$ to be defined below, and one expects that at  scales smaller than $L_{oz}$, isotropy and classical turbulent scaling recover.

Gravity waves (and inertia-gravity waves when rotation is included) are essential to understand the dynamics of large scales, the very scales where most of the energy resides. Under the assumption of stationarity, weak nonlinearities, weak dissipation and negligible forcing, one then obtains a balance which involves the pressure gradient and the gravity (together with the Coriolis force in the presence of rotation). Such a concept of balance has proven useful time and again in meteorology and oceanography, and many variants and issues have been considered to study the dynamics of synoptic scales. However, gravity waves couple nonlinearly on slow time scales,  and undergo wave steepening and breaking \cite{staquet_rev_02}, through resonant interactions as already described early on through the weak turbulence formalism \cite{bretherton, mccomas}. This leads for example to frontogenesis \cite{majda_96, mcwilliams_10}, as observed in the upper atmosphere or in the oceans. Overturning is also present in direct numerical simulations (DNS) of stratified turbulence at high resolution and high Reynolds number \cite{kimura, almalkie_12}. Thus, stratified turbulence is known to play a role in the vertical mixing of deep layers in the oceans, as observed for example in the central Pacific ocean \cite{wunsch_rev}, as well as in river bends and estuaries \cite{maccready}.

On the other hand, the role of helicity, the correlation between the velocity field ${\bf u}$ and its curl, the vorticity $\mathbf{\omega}$, has not received much attention outside the realm of astrophysics when considering the growth of large-scale magnetic fields \cite{branden_rev}. It is estimated routinely in the atmosphere, in conjunction with CAPE (convective available potential energy) in order to gauge the possibility of supercell convective storms to become strong \cite{moli}, and it may be a factor to take into account in the formation of hurricanes \cite{montgo} (see also \cite{marko, xU_03}). Taking the curl of the geostrophic balance (or in the absence of rotation, the so-called cyclostrophic balance) eliminates the pressure term, and thus also the thermodynamics that could enter through pressure gradient coupling with density variations and, for example, moisture. Thus, considering the dynamical role of helicity in stratified turbulence may allow for a focus on the dynamics of the atmospheric wind or oceanic current through decoupling from the pressure field. Note that cyclostrophic balance implies curvature of trajectories that is not due to rotation (which is neglected here), as is also the case at low latitudes or at small scales, for example in tornadoes \cite{winn}.

{Specifically, in this work we consider the effect of helicity in freely decaying stably stratified turbulence using direct numerical simulations. We show that helicity undergoes a very slow decay (when compared with the energy), and that cyclostrophic balance at large scales can play an important role in determining the decay of the flow.}

{In Sec. \ref{sec:method} we recall the incompressible Boussinesq equations for a stably stratified flow and we list some characteristic length scales, dimensionless number, and energies, relevant to the problem under consideration. We shortly present the numerical method used to integrate the equations, and we describe the different initial conditions imposed, reporting, in Table \ref{simulations}, the initial parameters chosen for our calculations. In Sec. \ref{sec:numerics} we mainly discuss the behavior of helicity for different initial conditions, we relate it to the velocity distribution in the horizontal and vertical directions and we interpret these results proposing a type of large-scale dissipative ``cyclostrophic'' balance. In what follows we study how the combined effects of different residual helicity values, different initial conditions and different Froude and Reynolds numbers affect the temporal decay of the total energy and the Fourier spectra for the total and potential energy and for the absolute value of helicity. Finally our results are summarized in \ref{sec:conclusion}.}

\section{Equations and methodology} \label{sec:method}
\subsection{Mathematical model and relevant parameters}
Stably stratified turbulence is studied here by means of the incompressible Boussinesq equations. In the presence of gravity $g$ and in the absence of forcing, with $\theta$ being the potential temperature fluctuations in units of velocity, the equations read:
\begin{eqnarray} 
\partial_t {\mathbf u} +{\mathbf u} \cdot \nabla {\mathbf u}  &=&  -\nabla P - N \theta\  e_z + \nu \Delta {\mathbf u}  \ , \\       \label{eq:mom}
\partial _t \theta\  +{\mathbf u} \cdot \nabla \theta\  &=& N w + \kappa \Delta \theta\  \ , \\  \label{eq:temp}
 \nabla \cdot {\bf u} &=&0 \ ;
\end{eqnarray} 
\noindent   $w$ is the vertical ($z$) component of the velocity in the direction of gravity, $P$ is the pressure, $\nu$ the viscosity, and $\kappa$ the diffusivity; in this paper we take a unit Prandtl number, $Pr=\nu/\kappa=1$. The square Brunt-V\"ais\"al\"a frequency is given by $N^2=-(g/\theta\ ) (d\bar \theta\ /dz)$, where $d\bar \theta\ /dz$ is the imposed background stratification, which is assumed to be linear. No filtering on the small scales ({\it e.g.}, hyperviscosity) is applied, a normal Laplacian being used for diffusion. 

In the ideal case (with $\nu=\kappa=0$), the Boussinesq equations conserve the total (kinetic plus potential) 
energy, 
$$ \frac{1}{2} \left<|{\bf u}|^2 + \theta\ ^2\right>=E_V+E_P \ , $$
and the point-wise potential vorticity 
$$PV= -N\omega_z +  \omega \cdot \nabla \theta\ , $$
which is a material invariant. For strong waves (or weak nonlinearities), $N>>1$,  and the linearized version of PV  leads to a quadratic invariant, $\left<PV_L^2\right>$, which is preserved by the Fourier truncation of the numerical algorithm and which is proportional to the so-called vertical enstrophy, $\left<\omega_z^2\right>$.
 
The volume integrated helicity, 
$$H = \left<{\mathbf u} \cdot {\mathbf \omega} \right>$$
is a topological quantity that characterizes the amount of ``twist'' and linkage in the flow \cite{tsinober}. In the 3D Euler equations, in the absence of stratification and of dissipation, helicity is an invariant, as is the kinetic energy, but in the above equations it is not. We nevertheless retain this diagnostic in order to make connections to isotropic and homogeneous turbulence, and to rotating flows for which helicity is invariant as well.

Relative helicity $\sigma_V$ measures the degree of alignment between velocity and vorticity, and as such, it can be viewed as a proxy measure of the amount of non-linearity in the flow, since $\sigma_V\sim \pm 1$ implies a Lamb vector $-{\bf u} \cdot \nabla {\bf u} + \nabla[u^2/2]={\bf u} \times \mathbf{\omega} \approx 0$. It can be defined as 
\begin{equation}
\sigma_V = \frac{H}{\sqrt{E_V\, Z_V} }, \label{rel_H}
\end{equation}
where $Z_V =  \left<\omega^2 \right>$ is the so-called kinetic enstrophy, proportional to the kinetic energy dissipation when viscosity is restored. One can define similarly a potential enstrophy $Z_P =\left<|\nabla \theta | ^2 \right>$, associated with the dissipation of potential energy $E_P$ when $\kappa\not= 0$.

Three characteristic length scales can be identified for this problem: ({\it i}) the Kolmogorov length scale $\eta$, where dissipation prevails for a Kolmogorov spectrum, ({\it ii}) the Ozmidov length scale, $L_{oz}$, and ({\it iii}) the buoyancy length scale $L_{b}$. For the initial conditions, these length scales are defined respectively as
$$\eta=\left(\frac{\nu^3 L_{0}}{U_{0}^3}\right)^{1/4} \ , \ L_{oz}=\frac{U_0^{3/2}}{N^{3/2}} \sqrt{\frac{1}{L_{0}}} \ , $$ $$ L_{b}=\frac{2\pi U_0}{N} \ ,$$
where $U_{0}$ is the initial $rms$ velocity, $L_0=2\pi/k_0$ is the initial integral scale, $k_{0}$ being the wave number where the initial excitation is centered, and $\epsilon_V\equiv dE_V/dt \approx U_0^3/L_0$ has been used as an estimate of the kinetic energy dissipation rate (under the assumption that the Ozmidov scale is resolved).

Three relevant dimensionless parameters can also be identified: ({\it i}) the Reynolds number $Re=U_0L_0/\nu$, ({\it ii}) the above mentioned Froude number, $Fr=U_0/(L_0N)$, and ({\it iii}) the buoyancy Reynolds number ${\cal R}=ReFr^2$.

{We finally define the so called reduced spectra $E$ as a function of the wavenumbers $k$, $k_{\perp}$ and $k_{\parallel}$; they are respectively $k=|\bf k|$, $k_{\perp}=|{\bf k_{\perp}}|=|{\bf k}| \sin \Theta$, with $\Theta$ the co-latitude in Fourier space with respect to the vertical axis of unit vector $\bf \hat{z}$, while $k_{\parallel}$ refers to the component of $\bf k$ in the $z$ direction.
If $U_{ij}(\bf k)$ is the velocity autocorrelation function  in Fourier space, we name its trace $U(\bf k)$. Under the assumption of homogeneity we define the axisymmetric energy spectrum as
 \begin{equation}
e(\mathbf{|k_{\perp}|}, k_{\parallel})=\int U({\bf k}) |{\bf k}| \sin \Theta d\phi=e(\mathbf{|k|}, \Theta),
\end{equation}
where $\phi$ is the longitude with respect to the $x$ axis; then we can define \cite{3072}
\begin{eqnarray}
E(k_{\perp})&=&\int e(\mathbf{|k_{\perp}|}, k_{\parallel})dk_{\parallel} \label{E1} , \\ E(k_{\parallel})&=&\int e(\mathbf{|k_{\perp}|}, k_{\parallel})dk_{\perp} \label{E2} , \\
E(k)&=&\int e(\mathbf{|k|},\Theta)|{\bf k}|d\Theta \label{E3} .
\end{eqnarray}
Similar definitions hold for $h(|{\bf k}|,\Theta)$, the axisymmetric helicity spectrum which is based on the antisymmetric part of the velocity correlation tensor \cite{3072}. From $h(|{\bf k}|,\Theta)$ we then derive the definitions for $H(k)$, $H(k_{\perp})$ and $H(k_{\parallel})$. 
Similar definitions hold as well for the potential energy distribution.
\subsection{Numerical model}
The numerical simulations are carried out using the Geophysical High-Order Suite for Turbulence (GHOST) code \cite{hybrid2011, gomez2005}. It is a pseudo-spectral framework available to the community and hosts a variety of partial differential equation (PDE) solvers optimized for studying turbulence in each of these PDE systems, in two-dimensional (2D) and three-dimensional (3D) geometries, for neutral and conducting fluids (MHD and Hall-MHD), and with solid-body rotation and stratification in the Boussinesq approximation. GHOST also  includes a passive scalar solver, as well as surface quasi-geostrophic and shallow water solvers, and several types of sub-grid scale models. In the 3D case, the grid is a cubic $[0,2\pi]^3$-periodic box, and with a $2^{nd}$--order explicit Runge-Kutta (RK) time stepping scheme; tests were conducted using a fourth order explicit RK scheme that showed neither qualitative nor quantitative differences when compared with the 
second order scheme at the resolutions used in this paper. De-aliasing is done by using a standard 2/3 rule. 
The importance of performing computations for this system in a cubic box with equal grid spacing in the horizontal and vertical directions was emphasized and demonstrated in \cite{waite2004}. We also 
use an explicit
time stepping method because of the necessity to resolve all time scales. For the code, a hybrid MPI/OpenMP parallelization methodology was developed \cite{hybrid2011}, with tests up to 98000+ compute cores on grids of up to 6144$^3$ points; the code  also has a third level of parallelization with the recent addition of GPU/accelerator support for the FFTs.
\subsection{Initial conditions} \label{subsec:database}
 
\begin{table*}
\centering
\begin{tabular}{|l|c|c|c|c|c|c|c|c|c|}
\hline
{\bf Initial v}  & {\bf $k_{0}$} &  $\nu$ & $Re$ & $Fr$ & ${\cal R}$ & $\eta$& $L_{oz}$ & $L_{b}$ & $\Delta$ \\
\hline
1. TG & 2-3 & 2.4e-04 & $\approx$ 3000    & 0.022 & 1.452 & 0.0041 & 0.0083 & 0.35 & 0.0246\\
2. TG$Fr/2$ & 2-3 & 2.4e-04 & $\approx$ 3000   & 0.011  & 0.363 & 0.0041 & 0.0074 & 0.175 & 0.0246 \\ 
\hline
3. ABC2C & 3-4 & 3.0e-04 &  $\approx$ 3000 & 0.022 & 1.452 & 0.0044 & 0.0059 & 0.25 & 0.0246 \\
4. ABC2C$2Re$ & 3-4 & 1.5e-04 &  $\approx$ 6000 & 0.022 & 2.904 & 0.0026 & 0.0059 & 0.25 & 0.0123 \\
5. ABC2C$2Fr$ & 3-4 & 1.5e-04 &  $\approx$ 6000 & 0.044 & 11.616 & 0.0026 & 0.0168 & 0.50 & 0.0123 \\
\hline
6. ABC$N0$ & 3-4 & 3.0e-04 &  $\approx$ 3000 & $\infty$ &$\infty$ & 0.0044 & $\infty$ &$\infty$ & 0.0123 \\
7. ABC & 3-4 & 3.0e-04 &  $\approx$ 3000 & 0.022 & 1.452 &0.0044 & 0.0059 & 0.25 & 0.0246 \\
\hline
8. ABC$2Re$ & 3-4 & 1.5e-04 &  $\approx$ 6000 & 0.022 & 2.904 & 0.0026 & 0.0059 & 0.25 & 0.0123 \\
9. ABC$2Fr$ & 3-4 & 1.5e-04 &  $\approx$ 6000 & 0.044 & 11.616 & 0.0026 & 0.0168 & 0.5 & 0.0123 \\
10. ABC$4Fr$ & 3-4 & 3.0e-04 &  $\approx$ 6000 & 0.088 & 46.464 & 0.0026 & 0.0474 & 1.0 & 0.0246 \\
11. ABC$Fr/2$ & 3-4 & 3.0e-04 &  $\approx$ 3000 & 0.011 & 0.363 & 0.0044 & 0.0021 & 0.125 & 0.0246 \\
\hline
12. RND$N0$  & 3-4 & 3.0e-04 &  $\approx$ 3000 & $\infty$ &$\infty$& 0.0044 & $\infty$ &$\infty$ & 0.0123 \\
13. RND  & 3-4 & 3.0e-04 &  $\approx$ 3000 & 0.022 & 1.452 & 0.0044 & 0.0059 &0.25 & 0.0246 \\
\hline
14. RND$2Re$  & 3-4 & 1.5e-04 &  $\approx$ 6000 & 0.022 & 2.904 & 0.0026 & 0.0059 & 0.25 & 0.0123 \\
15. RND$2Fr$  & 3-4 & 1.5e-04 &  $\approx$ 6000 & 0.044 & 11.616 & 0.0026 & 0.0168 & 0.50 & 0.0123 \\
16. RND$4Fr$  & 3-4 & 1.5e-04 &  $\approx$ 6000 & 0.088 & 46.464 & 0.0026 & 0.0474 & 1.0 & 0.0123 \\
17. RND$k2$  & 2 & 5.24e-04 &  $\approx$ 3000 & 0.022 & 1.452 & 0.0078 & 0.0104 & 0.437 & 0.0246 \\
\hline
\end{tabular}
\caption{List of the runs, computed on grids with either $n$=256 or $n$= 512. {The name of the runs (``Initial {\bf v}'') indicates the initial velocity field, and also summarizes some important properties (e.g., whether the Reynolds or Froude numbers were changed with respect to other runs with similar initial conditions); see the text for more details.} All these calculations have $U_{0}=0.5$, and dynamical parameters are evaluated at t=0;
  $k_0$ is the wavenumber range of the initial condition, $\nu$ is the viscosity and $Re$ the Reynolds number; Fr is the Froude number, ${\cal R}$ the buoyancy Reynolds number, $\eta$ the Kolmogorov (dissipative) length scale, $L_{oz}$ and $L_b$ are the Ozmidov and buoyancy length-scales, and $\Delta=2\pi/(n-1)$ is the grid resolution. Note that in general the Ozmidov length scale is not resolved, except for runs 5, 9 and 14, and that the buoyancy Reynolds number is of order unity except for the same three runs.}
\label{simulations} \end{table*}
 
A variety of runs were conducted at moderate Reynolds number in a 3D cubic domain, with isotropic discretization of $nx=ny=nz=n=256$, or $n=512$ points. The database of the runs is presented in Table \ref{simulations}, with parameters characterizing the initial conditions we employed.  Other runs were performed that confirm the conclusions of this work but which, for the sake of brevity, are not described in detail.

Three sets of initial velocities were investigated: Taylor-Green (TG), Arnold-Beltrami-Childress (ABC), and 
random isotropic. The TG velocity field is given by:
\begin{eqnarray}
v_x^{TG} &=& v_0^{TG} \sin (k_0x) \ \cos (k_0y) \ \cos (k_0z) \nonumber  \ , \\
 v_y^{TG} &=& -v_0^{TG} \cos (k_0x) \ \sin (k_0y) \ \cos (k_0z) \nonumber  \ , \\
  v_z^{TG} &=& 0 \ , \label{TGflow}
\end{eqnarray}
 and is globally non-helical ($\sigma_V \equiv 0$ at $t=0$).

The ABC initial condition for the velocity is specified as:
\begin{eqnarray}
v_x^{ABC} &=& v_0^{ABC} \left[B \cos(k_0 y) + C \sin(k_0 z) \right] \nonumber  \ , \\
v_y^{ABC} &=& v_0^{ABC}  \left[C \cos(k_0 z) + A \sin(k_0 x)  \right] \nonumber \ ,  \\ 
v_z^{ABC} &=& v_0^{ABC} \left[A \cos(k_0 x) + B \sin(k_0 y) \right] \ . 
\label{ABCflow}
\end{eqnarray}
As a superposition of Beltrami vortices (for which ${\bf u}= \pm  \mathbf{\omega}/k_0$), it is fully helical. Another approximately Beltrami initial condition was taken, namely the ${\bf v}^{ABC}$ written above with 
$v_z^{ABC}\equiv 0$ at $t=0$, in order to be able to compare with the evolution of the TG flow which also has $v_z=0$ initially. This second helical initial condition is called below ABC2C (2 components). The simulations were started using a superposition of these flows (either TG, or ABC, or ABC2C) for the wavenumber interval reported in Table \ref{simulations}.

We also examined random isotropic initial conditions (RND), which  use the method described in \cite{patterson} to generate a helical flow; it does so by correlating the velocity and vorticity in such a way that a specific amount of relative helicity can be chosen at $t=0$.

In the expressions (\ref{TGflow}) and (\ref{ABCflow}), we choose $v_{0}^{TG}=v_{0}^{ABC}=1$, and, for all the flows, the components of the  initial velocity are multiplied by a factor $f_0$, such that  the initial kinetic energy per unit volume is normalized to $U_0^2$:
\begin{equation}
\frac{f_0^{2}}{V} \left[\int_{V}(v_{x}^2+v_{y}^2+v_{z}^2)dV\right]=U_0^2 \ . \label{normaliz}
\end{equation}

Given Eq.~(\ref{normaliz}) and the definition of the relative helicity, Eq.~(\ref{rel_H}), the ABC flow has $\sigma_V(t=0) =1$, while the ABC2C flow displays $\sigma_V(t=0) =C^2/[C^2+(A^2+B^2)/2]$, which, for the choice $A=0.9$, $B=1$, $C=1.1$, gives $\sigma_V(t=0)\approx 0.57$. 
The choice of $A\neq B \neq C$ allows for a more rapid development of the flow and for turbulence to set in, by breaking more efficiently the symmetries present in the $A=B=C$ case.

The dimensionless Reynolds number $Re$, for the TG,  ABC and random flows, are defined as 
$$Re_{TG}=\frac{U_0 2\pi}{k_{0}\sqrt{3}\nu} \ , \ Re_{ABC, RND}=\frac{U_0 2\pi}{k_{0}\nu} \ . $$
The $\sqrt{3}$ factor in the definition for the TG flow comes from the fact that the individual Fourier shells are populated in different ways, for a given $k_0$, for the TG {\it vs.} ABC and random flows (e.g., \cite{sen2012}).
 
In each run the initial value of the potential temperature is $\theta\ (t=0) = 0$, so that buoyancy fluctuations 
 entirely develop from the initial velocity field.
Note that initial conditions need not be balanced since the time step for these runs is small enough to resolve the gravity waves that develop in each case.
A reason for using balanced initial conditions is related to the fact that one may want to investigate the development of imbalanced motions to start 
as a result of the turbulent mixing that develops, rather than by
choice of initial conditions. On the other hand, the non-linear terms will eventually develop these unbalanced motions as nonlinear coiling take place before and up to the peak of enstrophy.

\begin{table*} \centering \begin{tabular}{|l|c|c|c|c|c|}
\hline
{\bf Initial v} & $[E_{P}/E_{T}]^{\ast}$ & $\sigma_{V}^{0}$ & $\sigma_{V}^{\ast}$ &  $\sigma_{V}^{\ast\ast}$ \\
\hline
1. TG & 0.01 &  0.0    & 0.0 & 0.0 \\
\hline
6. ABC$N0$ & 0.00 & 0.99    & 0.13 & 0.103\\
7. ABC & 0.15 &   0.99 & 0.22 & 0.76 \\
\hline
12. RND$N0$  &  0  & 1.00  &  0.13 & 0.08\\
13. RND  & 0.18 &    1.00 & 0.02 & 0.11\\
\hline
15. RND$2Fr$  & 0.18 &    1.00 & 0.04 & 0.13\\
17. RND$k2$  & 0.17 &  1.00  &  0.11 & 0.70\\
\hline
\end{tabular}
\caption{Some relevant parameters for a selected set of the runs listed in Table \ref{simulations}. The ratio of potential to kinetic energy is given at peak of enstrophy $t^*$, while the relative helicity is reported at $t=0$ ($\sigma_V^0$), at the peak of the enstrophy ($\sigma_{V}^{\ast}$), and at $t=20$ ($\sigma_{V}^{\ast\ast}$). Note the increase of 
{relative helicity over time in all stratified cases, after an initial startup phase.}
} \label{peak_enstrophy} \end{table*}

\section{Results} \label{sec:numerics}
\begin{figure*}[h!tbp] \centering
\resizebox{7cm}{!}{\includegraphics{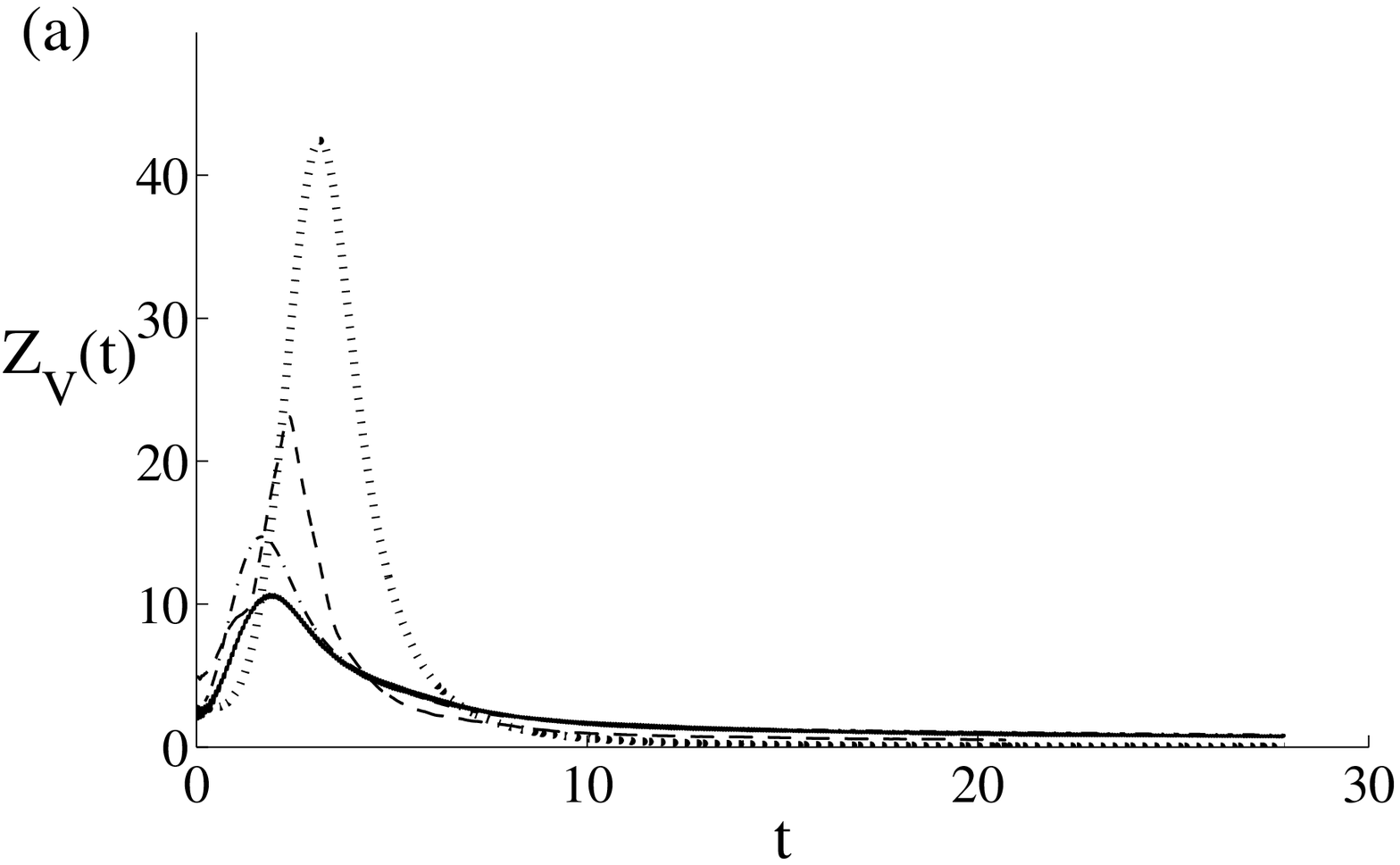}}
\resizebox{7cm}{!}{\includegraphics{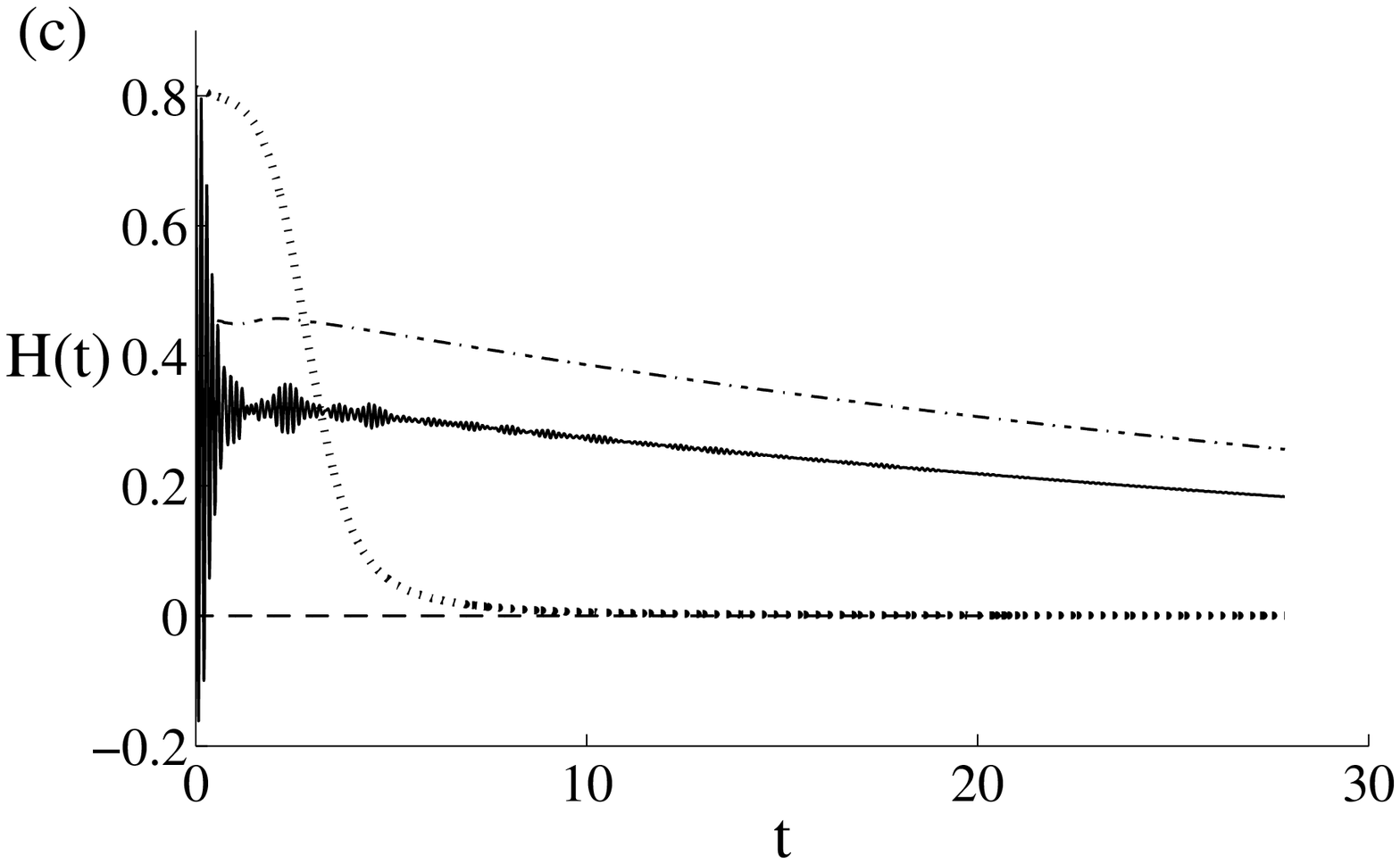}}
\resizebox{7cm}{!}{\includegraphics{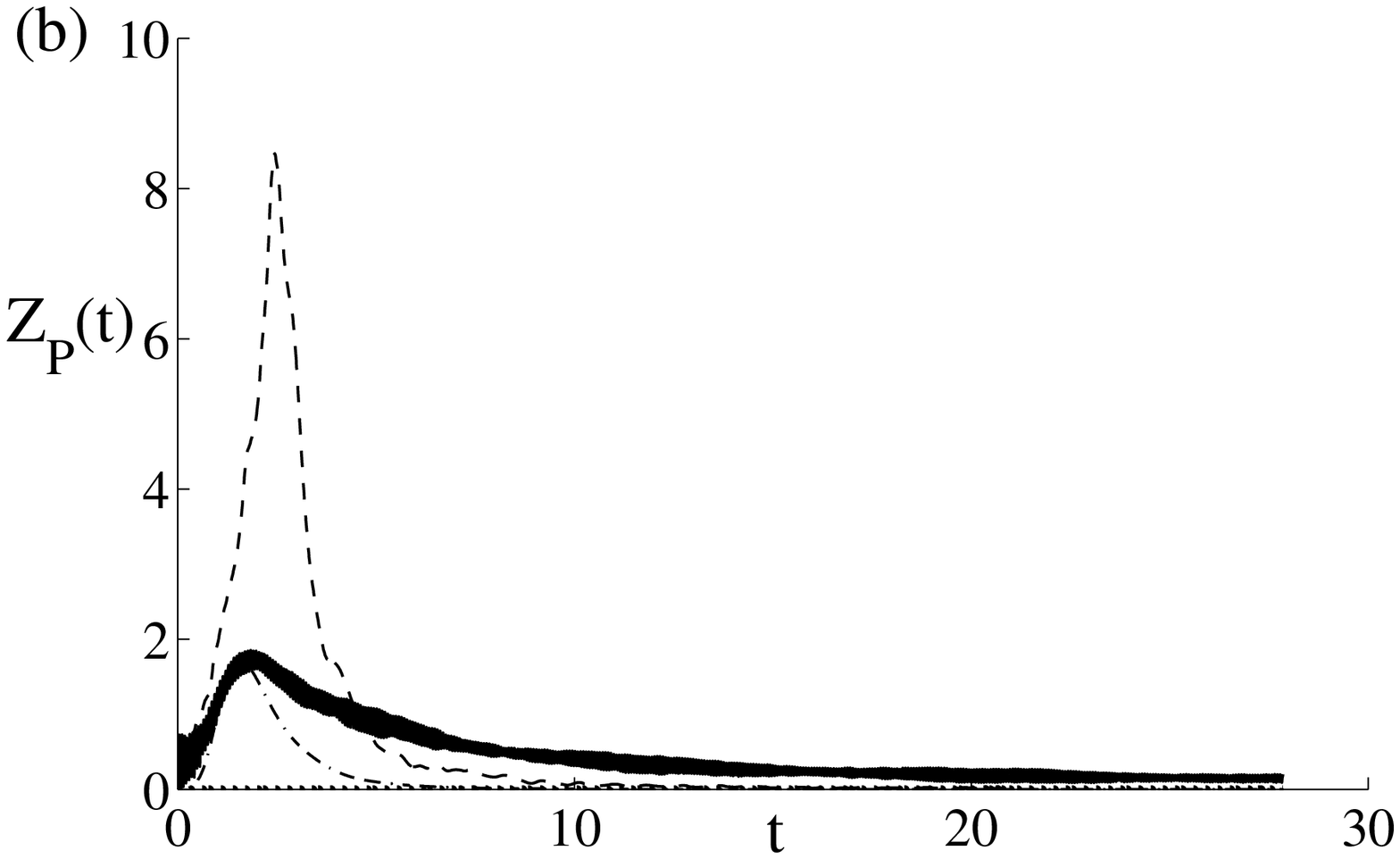}}
\resizebox{7cm}{!}{\includegraphics{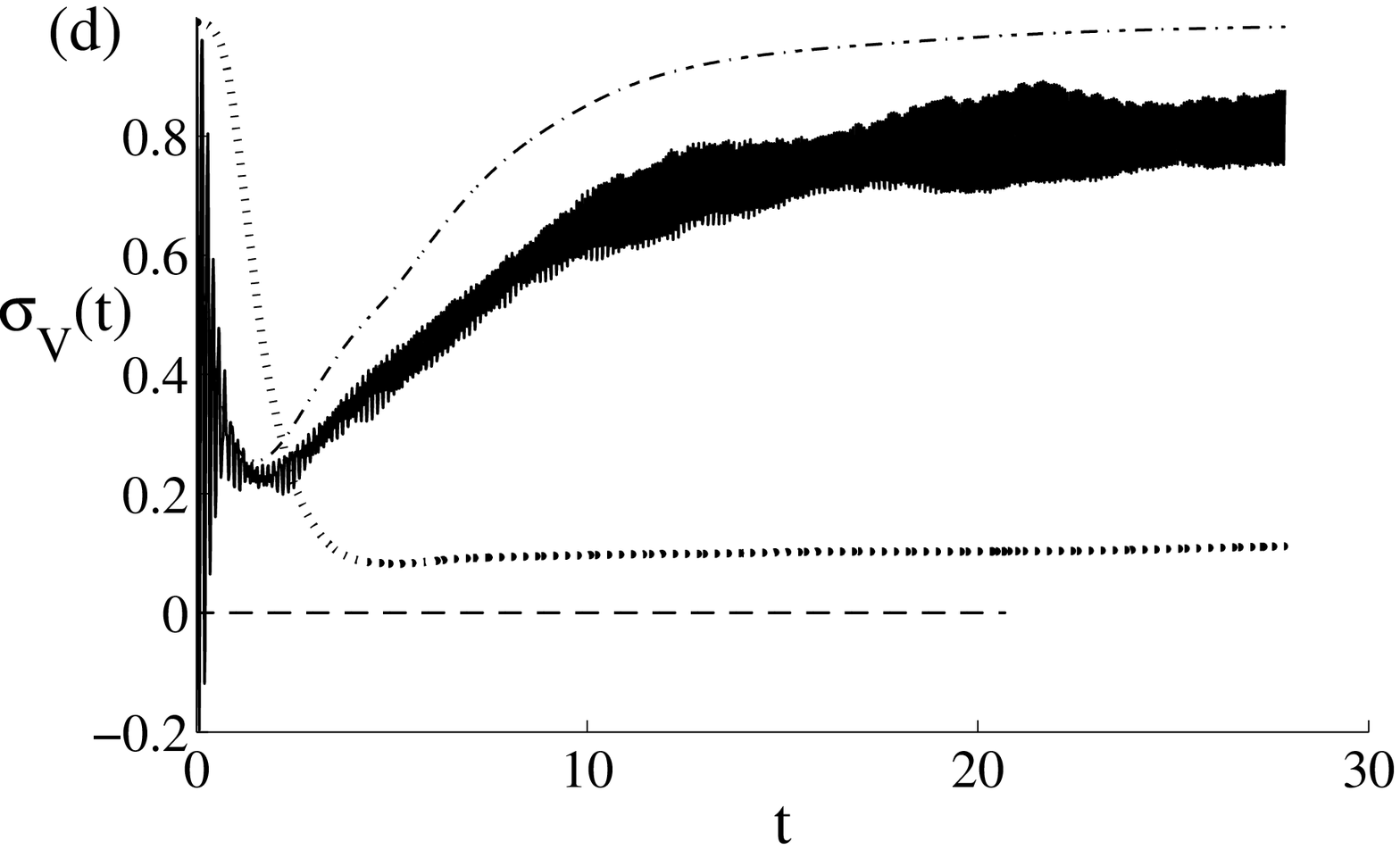}}
\caption{\label{TG_ABC_Ht}
Temporal evolution of the kinetic enstrophy (a), potential enstrophy (b), total helicity (c) and relative helicity (d), for several initial conditions, namely an unstratified ABC (run 6, ABC$N0$, dotted line), and three flows with $Fr \approx 0.022$: an ABC (run 7, ABC, solid line), an ABC2C (run 3, ABC2C, dash-dotted line), and a TG flow (run 1, TG, dashed line). The helicity of the TG flow, initially zero, remains negligible (dashed line), whereas for the two other  stratified and helical flows, the helicity decay is remarkably slow compared to the non-stratified case  (dotted line). Note the quasi-maximal helical state reached in the ABC2C case (dash-dotted), {especially clear in the evolution of $\sigma_V(t).$}
} \end{figure*} 
\subsection{The temporal decay of helicity} \label{subsec:helicity}
\begin{figure*}[h!tbp] \centering
\resizebox{6cm}{!}{\includegraphics{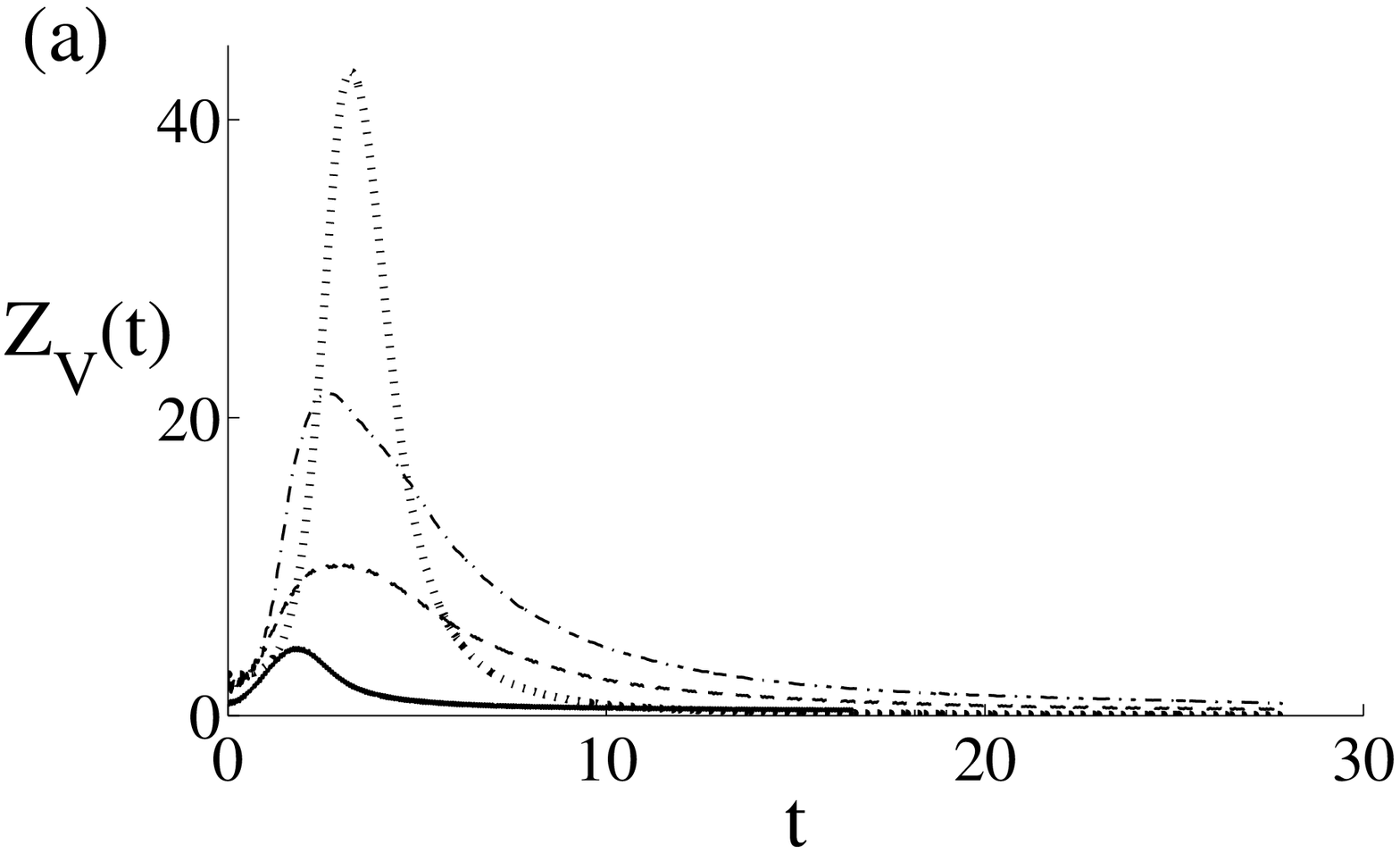}}
\resizebox{8cm}{!}{\includegraphics{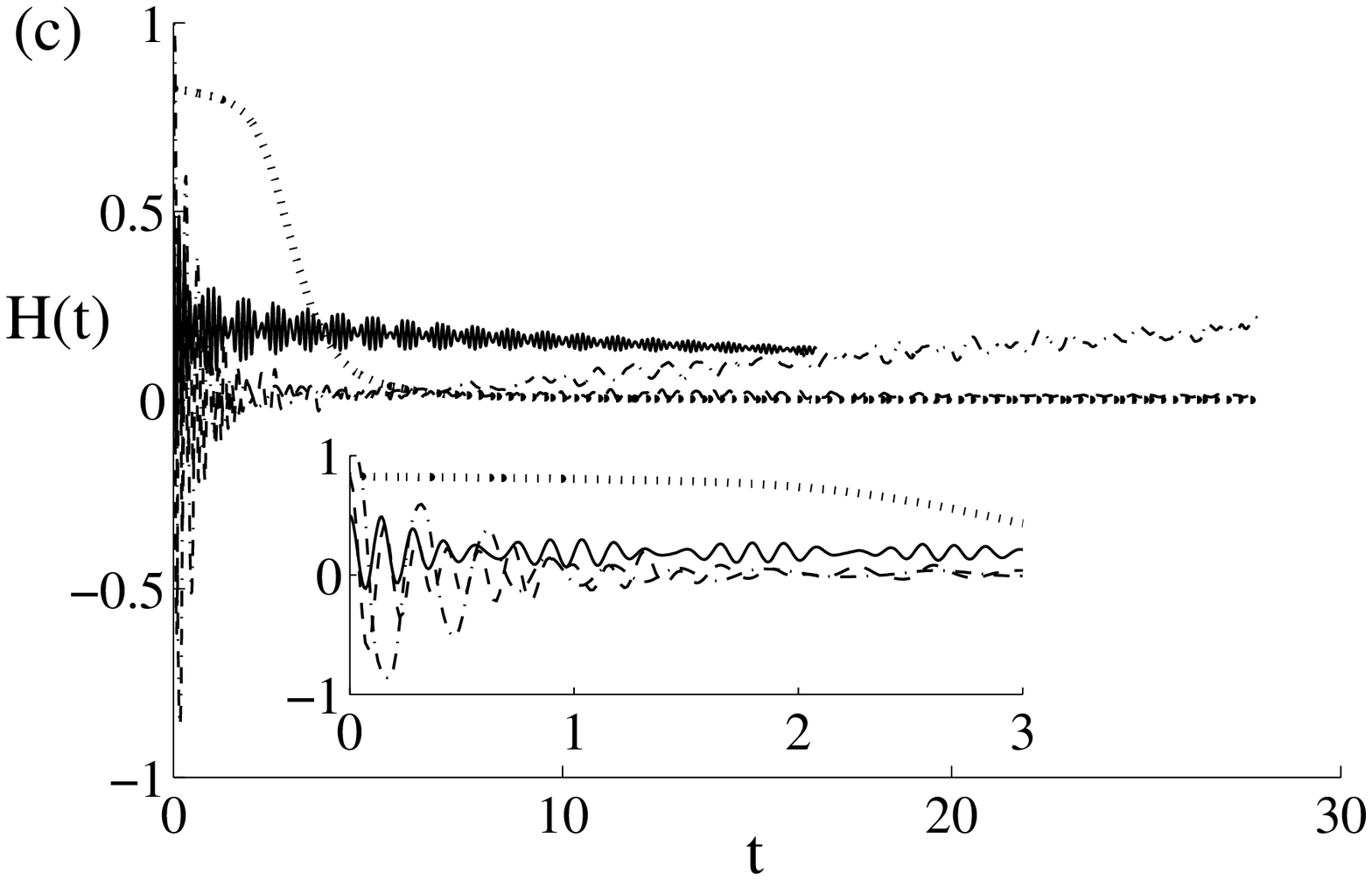}}
\resizebox{6cm}{!}{\includegraphics{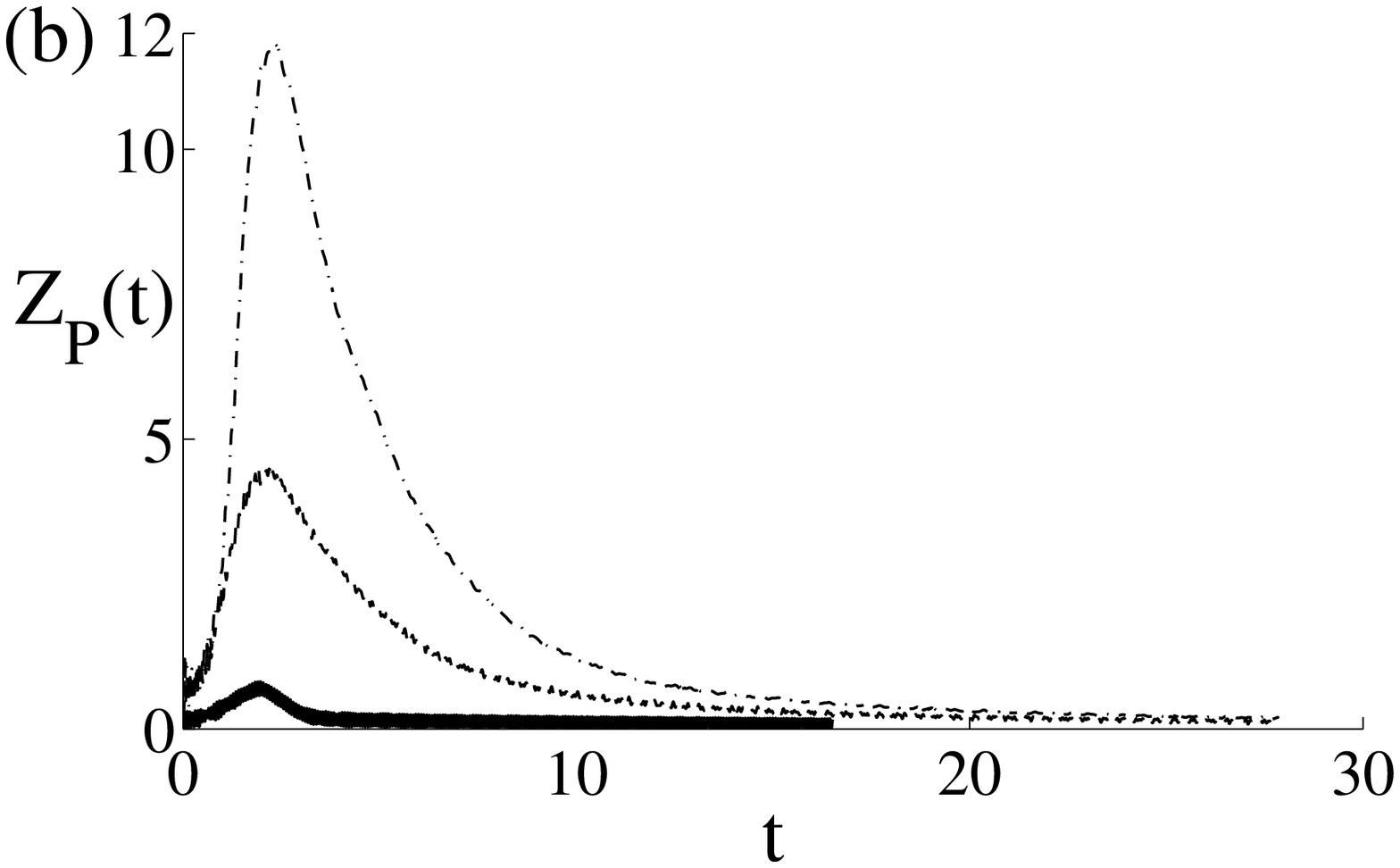}}
\resizebox{8cm}{!}{\includegraphics{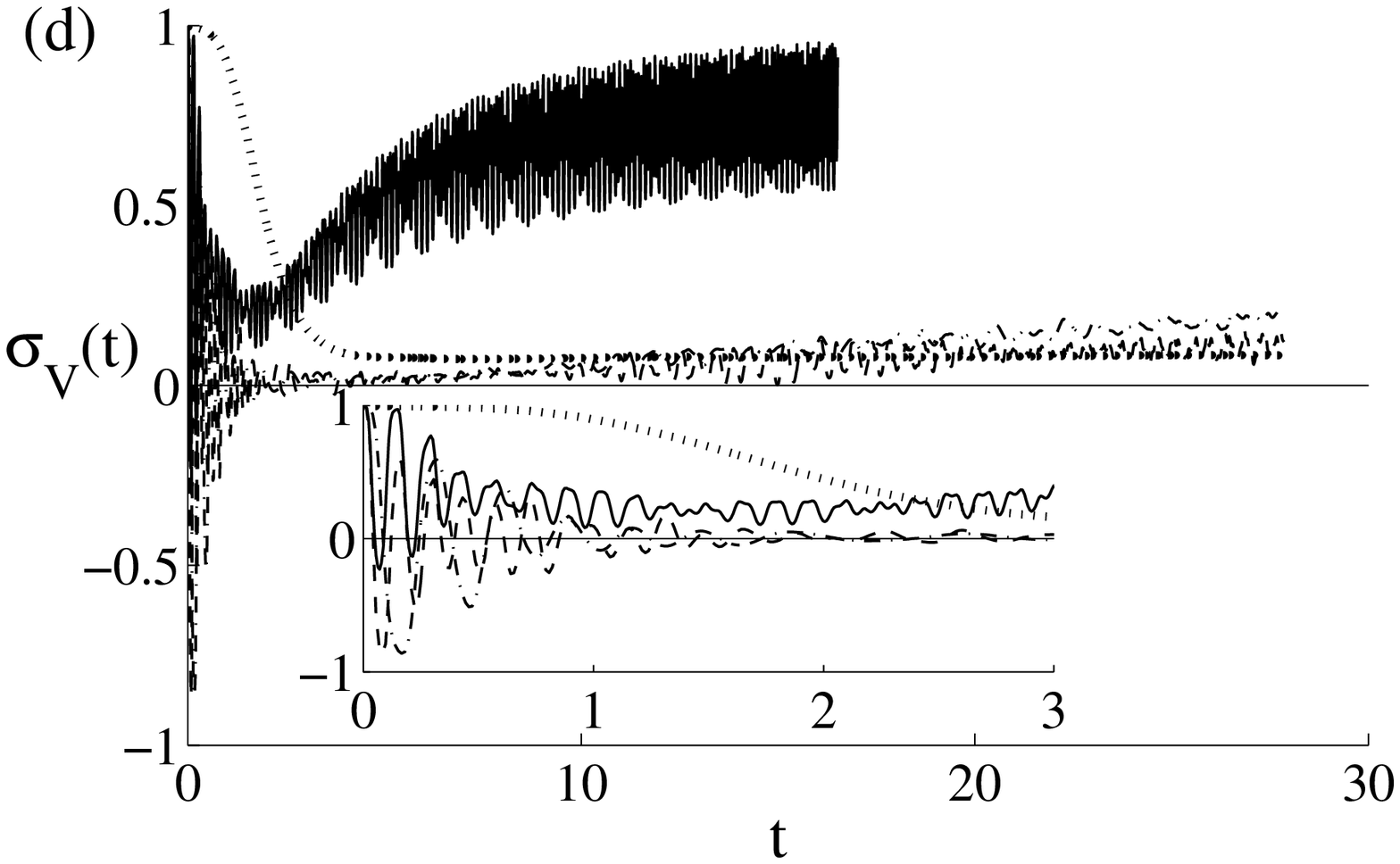}}
\caption{\label{RDM_k2_3_2F_Ht}Temporal evolution of the kinetic enstrophy (a), potential enstrophy (b), total helicity (c) and relative helicity (d),  for random initial conditions: an initially unstratified  flow (run 12, RND$N0$, dotted line), a flow with $Fr \approx 0.022$ (run 13, RND, dashed line), a flow with $k_{0}=2$ and $Fr \approx 0.022$ (run 17, RND$k2$, solid line), and a flow with $Re=6000$ and $Fr \approx 0.044$ (run 15, RND$2Fr$, dash-dotted). Note that all flows are close to maximum helicity initially (see Table \ref{simulations}). The insets in (c) and (d) give a zoom on the early-time evolution of helicity.
} \end{figure*} 
\begin{figure*}[h!tbp] \centering
\resizebox{7cm}{!}{\includegraphics{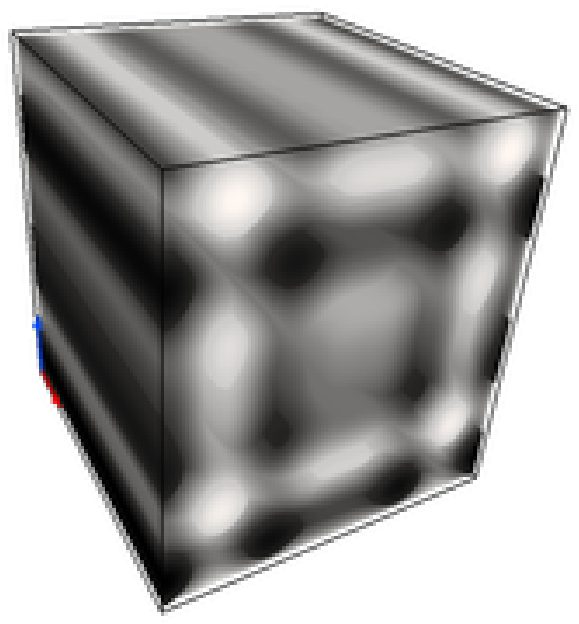}}
\resizebox{7cm}{!}{\includegraphics{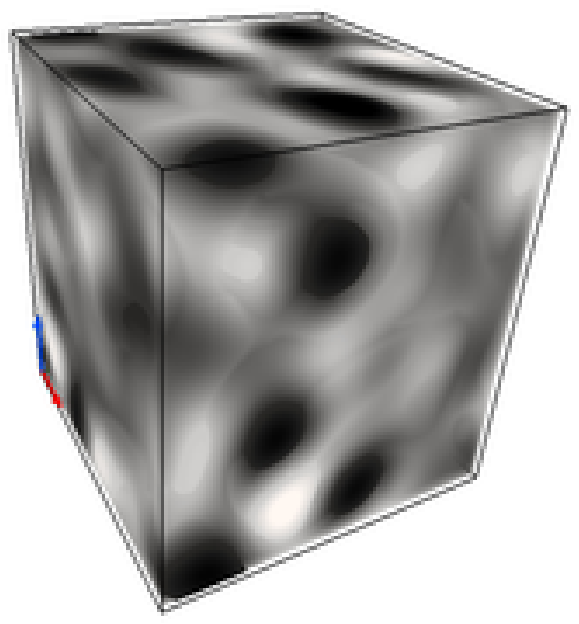}}
\resizebox{7cm}{!}{\includegraphics{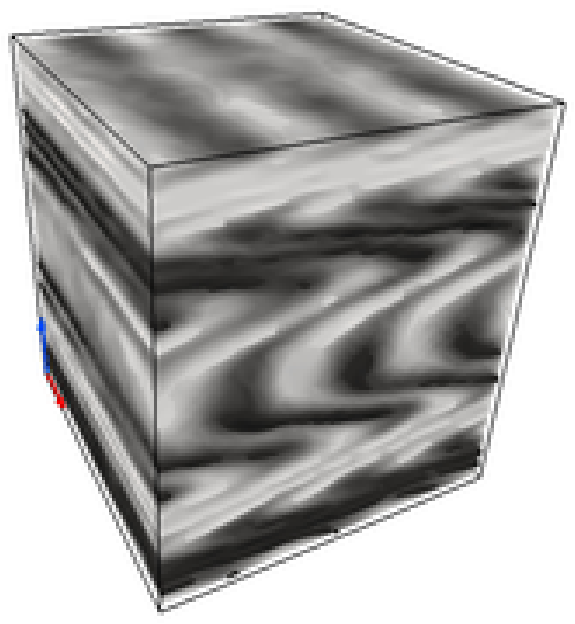}}
\resizebox{7cm}{!}{\includegraphics{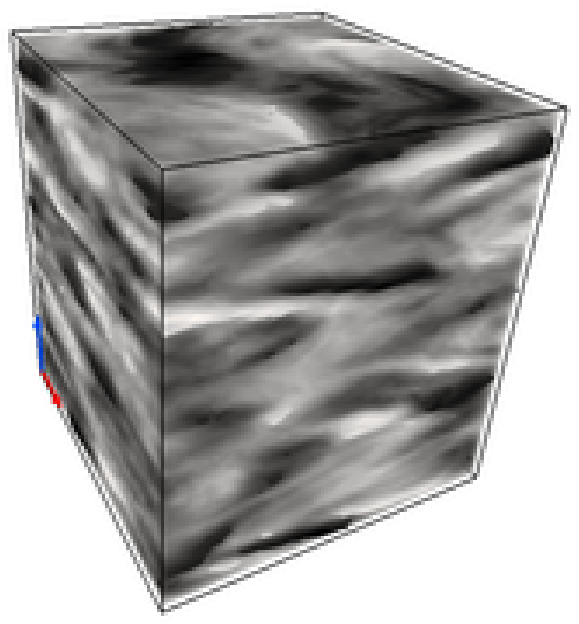}}
\resizebox{7cm}{!}{\includegraphics{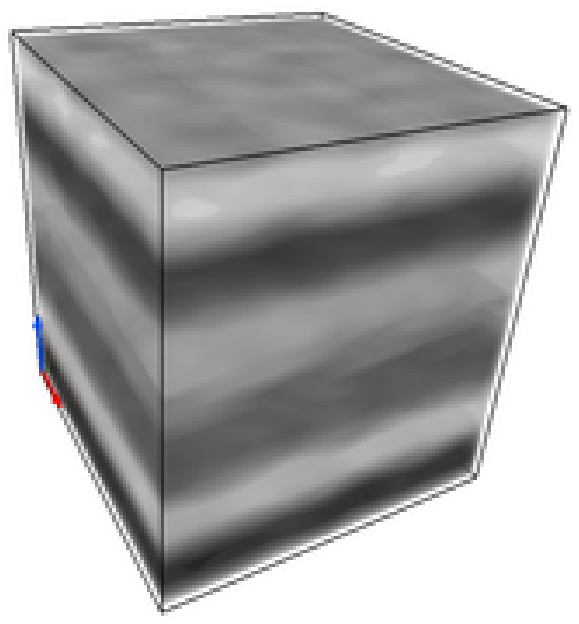}}
\resizebox{7cm}{!}{\includegraphics{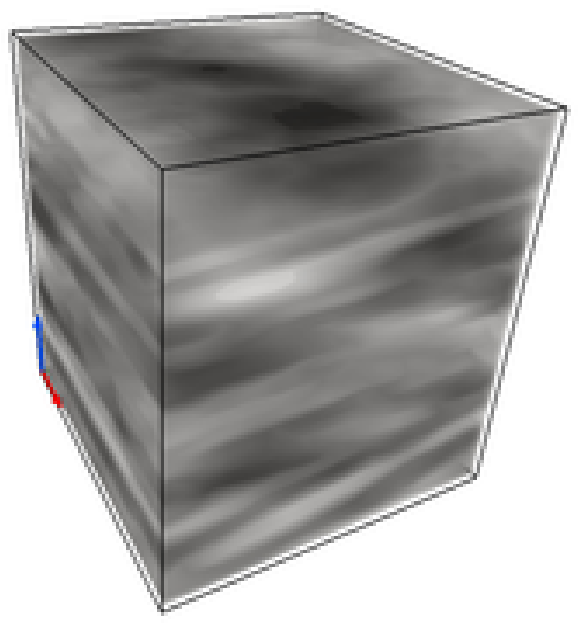}}
\caption{\label{VAPORvx}
(Color online) Distribution of the $x$ (horizontal) velocity $v_x$ for the ABC$2Re$ flow (left) and the RND$2Re$ flow (right) at times (from top to bottom) $t=0$, $t=t^*$ (i.e., at the peak of dissipation), and $t=20$. Light corresponds to low negative values, dark to high positive values. The vertical direction is indicated by the upward (blue) arrow. 
{Observe the strong shear layers that have developed by the peak in enstrophy, and which later on 
 begin to smooth out.}
} \end{figure*} 

\begin{figure*}[h!tbp] \centering
\resizebox{7cm}{!}{\includegraphics{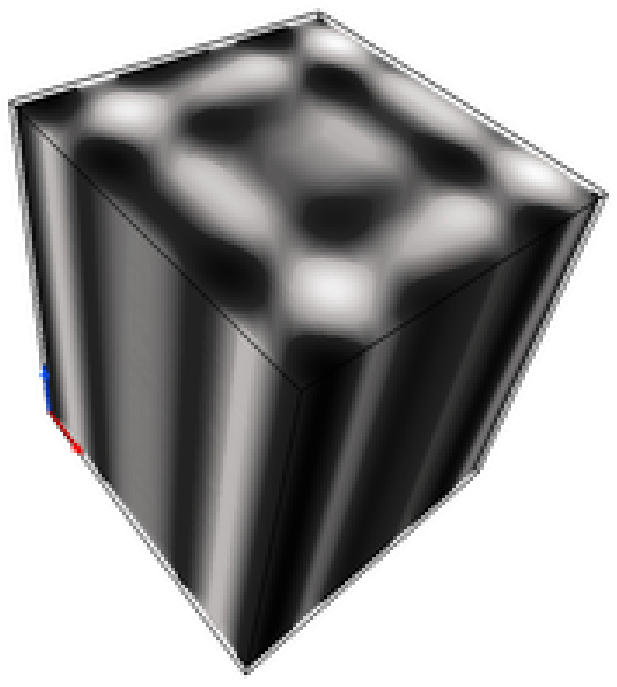}}
\resizebox{7cm}{!}{\includegraphics{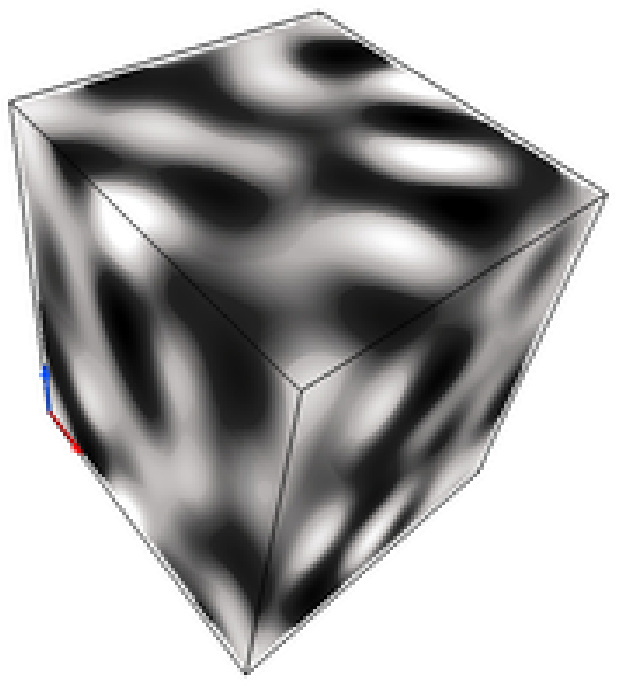}}
\resizebox{7cm}{!}{\includegraphics{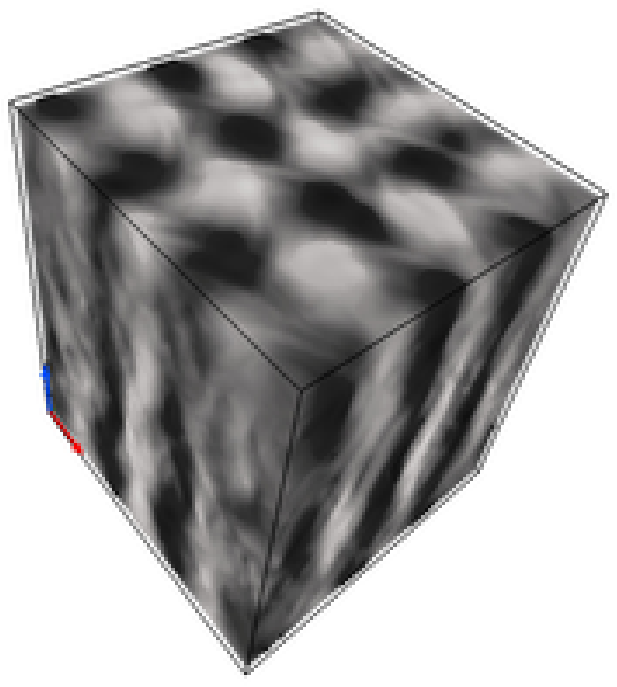}}
\resizebox{7cm}{!}{\includegraphics{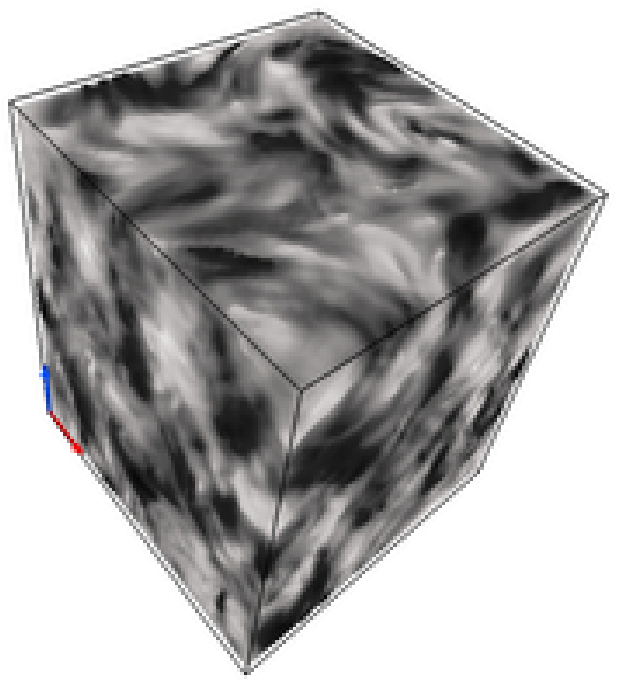}} 
\resizebox{7cm}{!}{\includegraphics{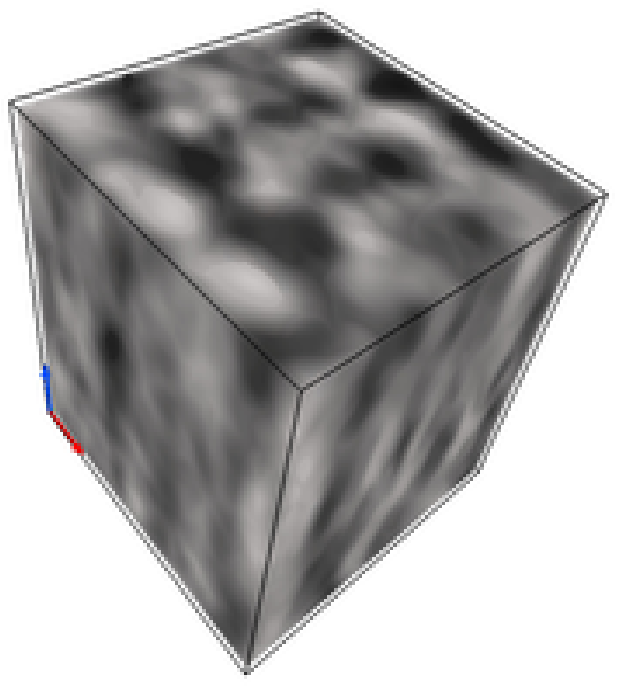}}
\resizebox{7cm}{!}{\includegraphics{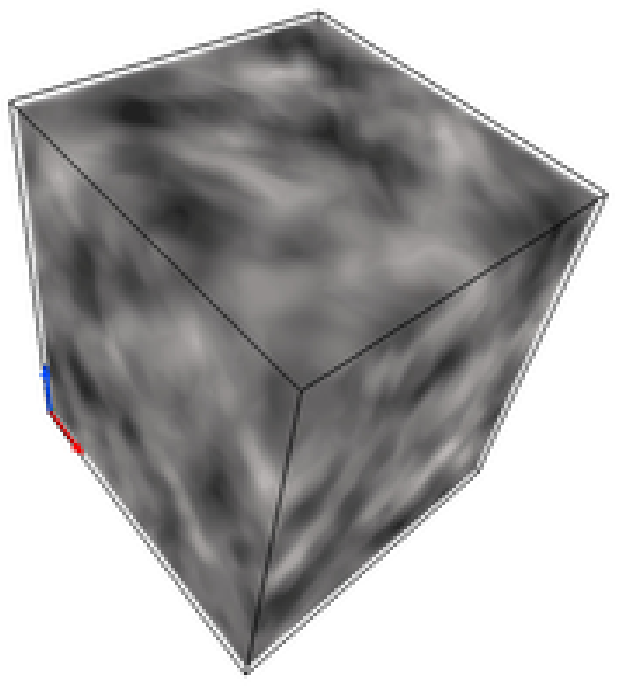}}
\caption{\label{VAPORvz}
(Color online)  Distribution of the vertical velocity $v_z$ for the ABC$2Re$ flow (left) and the RND$2Re$ flow (right) at times (from top to bottom) $t=0$, $t=t^*$, and $t=20$. Light corresponds to low negative values, dark to high positive values.
{Observe the more turbulent flow in the random case.}
} \end{figure*}

Here we examine the temporal behavior of the flows whose initial conditions are given in Table \ref{simulations}. For some of the simulations, the ratio between the potential energy and the total energy at the peak of the enstrophy, the value of the relative helicity at the initial time $\sigma_{V}^0$, at the peak of the enstrophy $\sigma_{V}^*$, and at $t=20$, $\sigma_{V}^{**}$, are reported in Table \ref{peak_enstrophy}.

{ A hint as to the outcome one may expect from the time evolution of helicity is obtained by  considering the dynamical equations. By taking the inner product of Eq. (1) with $\mathbf{\omega}$ and volume-averaging, it can be shown that the time derivative of the total helicity is
\begin{equation}
\frac{dH}{dt} = -2N\left<\theta\  \omega_z \right> -2\nu Z_H, \hskip0.1truein Z_H=\left<{\mathbf \omega} \cdot \nabla \times {\mathbf \omega} \right>, \ 
\label{eq:dHdt}
\end{equation}
with $Z_H$ the helical enstrophy (sometimes called super-helicity \cite{hide_02}), a pseudo-scalar as helicity itself. Note that, locally (as opposed to globally), helicity can be produced through alignment of vorticity and shear \cite{matthaeus}. {However, globally and for an initially helical flow, $Z_H$ is responsible for the viscous decay of helicity. In the absence of dissipation, helicity is conserved for non stratified flows, while the first term on the r.h.s.~ of Eq.~(\ref{eq:dHdt}) can act as a source or a sink for stratified flows, thus breaking the conservation.}}

In Fig.~\ref{TG_ABC_Ht}  (a) and (b) we first show the temporal evolution of the kinetic enstrophy, $Z_V(t)$, and potential enstrophy, $Z_P(t)$. All the flows have $Fr\approx 0.022$, with, as initial conditions, either an ABC flow, an ABC2C flow, or a Taylor-Green flow. For comparison, one case with ABC initial conditions, is unstratified. The time in all the figures is expressed in units of the initial eddy turn-over time $\tau_{NL}=2\pi/(k_{0}U_0)$}. We see that  the production of enstrophy, and therefore, the transfer of energy to small scales, is {\it damped} substantially in the presence of stratification, both for the ABC and TG initial conditions, although its maximum is not considerably {\it delayed}. This is expected  because the effect of waves through the buoyancy forces is to reduce the nonlinear interactions, as well as to suppress, in part, the vertical velocity component. Observe that the stratified TG flow displays a significantly stronger peak for the kinetic and potential enstrophy than the stratified ABC-like flows for which the nonlinear terms are (initially) equal to zero.

In Fig.~\ref{TG_ABC_Ht} (c) and (d) the temporal evolution of the total helicity and relative helicity for the same flows is shown. In the absence of stratification and for ABC initial conditions, helicity decays rather rapidly (exponentially after an initial nonlinear phase) and is close to zero for $t>6$. On the contrary, stratified ABC and ABC2C flows display slow decays of helicity, with strong oscillations at first when $v_z(t=0)\not= 0$, linked to gravity waves. Finally, we observe that the TG flow has zero initial helicity, and none is created by stratification.  {It is interesting that in the stratified cases helicity decays almost linearly with time, and much more slowly than the energy, which, as will be shown later and as is often the case for turbulent flows, decays as a power law with time.}

In the absence of stratification the relative helicity $\sigma_{V}(t)$ decays rapidly, the flow becoming closer to mirror-symmetric. On the other hand, when gravity is switched on and  initial helicity is non-zero, $\sigma_{V}(t)$ approaches the maximum value. This may be simply due to the fact that helicity decays more slowly than energy, leading to a growth of their ratio. More precisely, it can be seen from the data shown in Fig. \ref{TG_ABC_Ht} and later in Fig. \ref{TG_Etb} that $d_{t}H<<(d_{t}E_{V})^{1/2}(d_{t}Z_{V})^{1/2}$, hence, in Eq.~(\ref{rel_H}) the denominator increases much faster than the numerator, causing $\sigma_{V}$ to grow. 

The flows analyzed until now are well-ordered, centered at large scales and with phase coherence between modes at $t=0$. We also examined initial conditions with randomized phases yet maintaining a high initial relative helicity (see Table \ref{peak_enstrophy}). In Fig.~\ref{RDM_k2_3_2F_Ht} (a) and (b) the behavior of $Z_V(t)$ and $Z_P(t)$ for random flows with $k_{0}=3-4$ or $k_{0}=2$ is shown. It is seen that the unstratified flow displays a higher peak of the kinetic enstrophy, as noticed before, followed by the less stratified case ($Fr\approx 0.044$); while, when $Fr\approx 0.022$ the case with $k_{0}=2$ has a lower peak than the case with larger $k_{0}$. This is likely due to the fact that when $k_{0}=2$ the flow preserves some helicity, which inhibits the energy decay. A similar behavior is seen for the potential enstrophy: the peak is higher for the less stratified case, and is lower when helicity is better preserved.
 \begin{figure}[h!tbp] \centering
\resizebox{8cm}{!}{\includegraphics{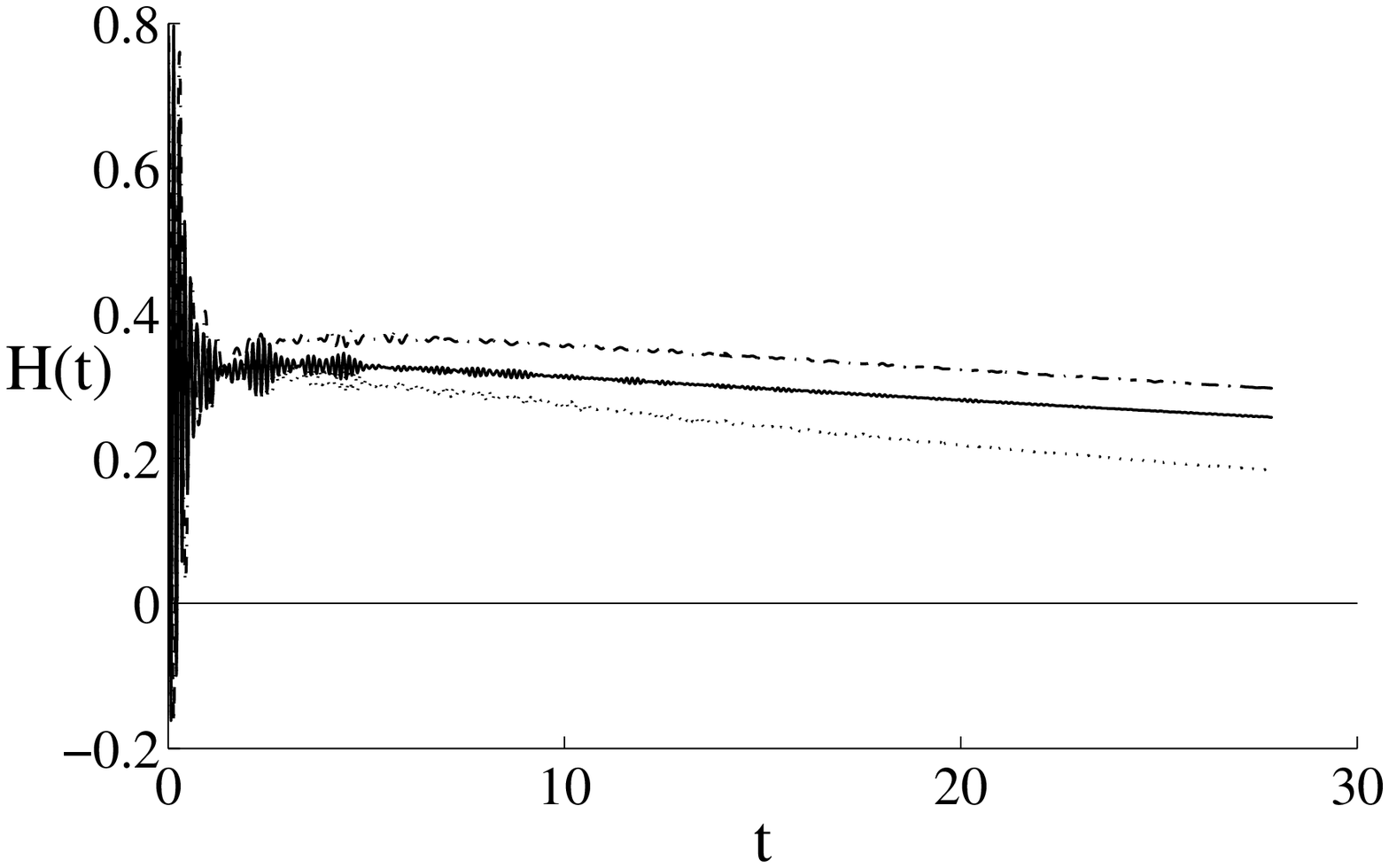}}
\caption{\label{ABCs_Ht}
Temporal evolution of the total helicity for  ABC flows with $Re=6000$ and $Fr=0.022$ (run 8, ABC$2Re$, solid line), $Re=6000$ and $Fr=0.044$ (run 9, ABC$2Fr$, dashed-dotted line), and $Re=3000$ and $Fr=0.022$ for reference (run 7, ABC, dots).}
\end{figure}

In random flows,  the time evolution  of the helicity and the relative helicity reveals a more complicated behavior than in ABC and TG flows. It is indeed observed that if the injection wave number is $k_{0}=2 $ (or $k_{0}=1$, not displayed), then, consistently with what we reported in the ABC cases, the relative helicity grows, while, for smaller-scale initial conditions, the relative helicity decays to a value close to zero (see Fig. \ref{RDM_k2_3_2F_Ht}). If however, the $Fr$ number is increased, a small growth of the relative helicity is observed also for the $k_{0}=3-4$ case. A similar run at twice the Reynolds number (using a grid with a resolution of  $512^3$ points) reveals no difference in the helicity behavior. The reason why random flows at $k_{0}=2$ behave similarly to ABC flows is likely due to the fact that their initial conditions resemble an ABC flow, especially  
with regard to the velocity distribution, with rather well-identified large scale structures; in the case of the ABC flow corresponding to three Beltrami waves, also called the Roberts flow, it resembles periodic cylindrical vortices (see \cite{dombre} for a visualization of the ABC flow at $t=0$). For higher initial $k_0$, the randomness (and the existence of more modes that can be excited in a given Fourier shell) destroys these well-organized structures.

Different snapshots of the $x$--  and $z$--components of the velocity, are given, respectively, in Fig. \ref{VAPORvx} and Fig.~\ref{VAPORvz} for the ABC2Re and RND2Re flows. Note that the horizontal cylindrical vortices, present in the ABC flow at $t=0$, are being sheared and develop into a zig-zag shape at the peak of dissipation, reminiscent of the zig-zag instability found in laboratory experiments \cite{chomaz}. However, this does not appear as being as well defined in the random runs. {At later times the $x$-component of the velocity is redistributed in horizontal layers of alternating sign; this structure is much better ordered for the ABC flow but is still visible for random flows.} It is clear from inspection of Fig.~\ref{VAPORvz} that the vertical velocity keeps its cylindrical structuring in the ABC2Re flow, whereas it becomes more turbulent in the case of the RND2Re flow. { It is also observed that, at a later stage, the ABC vertical columnar structures that survive alternate sign in time, suggesting a coupling with internal gravity waves.}

Finally, the effect of changing the Reynolds and Froude number on the time evolution of helicity for an ABC flow is shown in Fig. \ref{ABCs_Ht}. By increasing the value of these dimensionless numbers, the total helicity increases accordingly. The same phenomenon is observed, even if in smaller proportion,  in the case of random flows.
\subsection{Cyclostrophic balance} \label{subsec:balance}

By measuring separately the first and second term on the r.h.s.~of the $dH/dt$ expression we find that, at early times, characterized by high amplitude gravity waves, the first term 
dominates. After the peak of enstrophy, the two terms are small and balanced and cause  $dH/dt \approx 0$. For the ABC2C flow with $v_z(t=0)=0$, there is no need for radiation of gravity waves to adjust to large-scale balance and  the two terms are small and comparable at all times. This may explain the large values of residual helicity at late stages.
 
We observe that both the ABC and the random $k_{0}=2$ flows initially excite smaller amplitude oscillations of  the first term than the $k_0=3-4$ random flows do. These oscillations cease roughly at the peak of the enstrophy, when nonlinear coupling of waves takes over the linear phase. If the oscillation amplitude is large, $H(t)$ goes to zero even before the peak of  enstrophy (this is observed for  the $k_0=3-4$ random flows), an irremediable situation in the case of decaying flows, while, if the oscillation amplitude is small, $H(t)$ will have a non-zero residual value at the peak of enstrophy, and from there on it will  decay very slowly (ABC and $k_0=2$ random cases).

Increasing the Reynolds number does not seem to modify the behavior of $H(t)$ before the peak of the enstrophy, because the dynamical evolution in this first stage is plausibly wave-dominated. If the Froude number is varied instead, from $Fr=0.022$ to $Fr=0.044$, the initial oscillations are reduced and now the random case with $k_{0}=3-4$ also sees a growth of the relative helicity (see Fig.~\ref{RDM_k2_3_2F_Ht}). 
We have to conclude that the growth of relative helicity, or similarly the slow decay of helicity, is non-monotonic with Froude number since in the case $N=0$,  a decay of $\sigma_{V}(t)$ is observed. 
Another distinction which may help interpreting the different behavior of helicity in the ABC flow and the random flow, is that while in ABC flows helicity is organized in large scale coherent structures, in the structureless random flow it is not (except for $k_0=2$).
 
We are thus led to conjecture that, similarly to the geostrophic balance in rotating stratified turbulence, another large-scale balance, linked to dissipation, can take place in helical flows in the presence of stratification alone. Such {\it viscous cyclostrophic balance} would be consistent with $dH/dt \approx 0$, as observed for the Froude numbers considered. By neglecting the time derivative and the nonlinear term in Eq.~(1), taking the curl of the reduced equation and dotting it with velocity, we arrive at the following 
balance equation:
\begin{equation}
N{\mathbf u}\cdot \nabla \times (\theta\ e_z) = -\nu {\mathbf u}\cdot \nabla^2 {\mathbf \omega} 
= \nu {\mathbf u} \cdot \nabla \times \nabla \times  {\mathbf \omega} .
\label{eq:bal} 
\end{equation} 
Note that at this stage there is no space average being computed; however, upon space integration, this gives the relationship between vertical gradients of the temperature and the super helicity, as derived before,
 a balance that may hold at large scales where the waves dominate the dynamics. This balance cancels the two terms on the r.h.s.~of Eq.~(\ref{eq:dHdt}) thus explaining the slow decay of helicity when it is initially at large scales. What happens with this balance when the buoyancy Reynolds number is increased will require runs at higher resolution (or lower stratification) than what is considered in the present paper, and will be the topic of a subsequent study (see \cite{almalkie_12} for a large run where the Ozmidov scale is clearly resolved). However, it is interesting to note that viscous forces are known to play other important roles in stratified turbulence, such as when gravity waves break and alter the conservation of total energy and of potential vorticity (PV), leading to the creation of PV dipoles (see \cite{MCI_90}).
\subsection{Energy decay} \label{subsec:decay}
\begin{figure}[h!tbp] \centering
\resizebox{8cm}{!}{\includegraphics{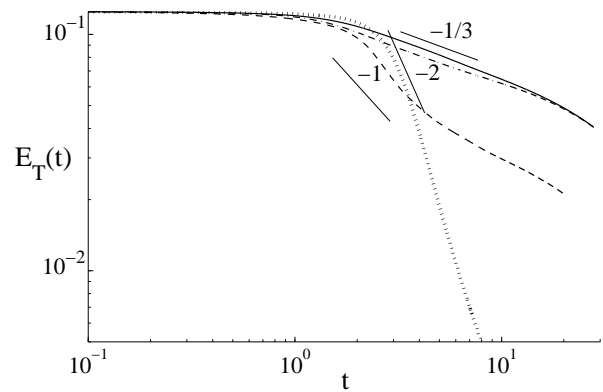}}
\caption{\label{TG_Etb} 
Temporal evolution of the total energy, with different scaling laws given as indications. ABC flow (run 7, ABC, solid line),  ABC2C flow (run 3, ABC2C, dash-dotted line), and TG flow (run 1, TG, dashed line), the latter two  having initially $v_z=0$. All runs have $Re \approx 3000$ and $Fr \approx 0.02$ except for the ABC flow with $N=0$ which is shown for comparison with a dotted line (run 6, ABC$N0$). Note the particularly slow decay for ABC runs 3 and 7.
} \end{figure}
\begin{figure*}[h!tbp] \centering
\resizebox{7cm}{!}{\includegraphics{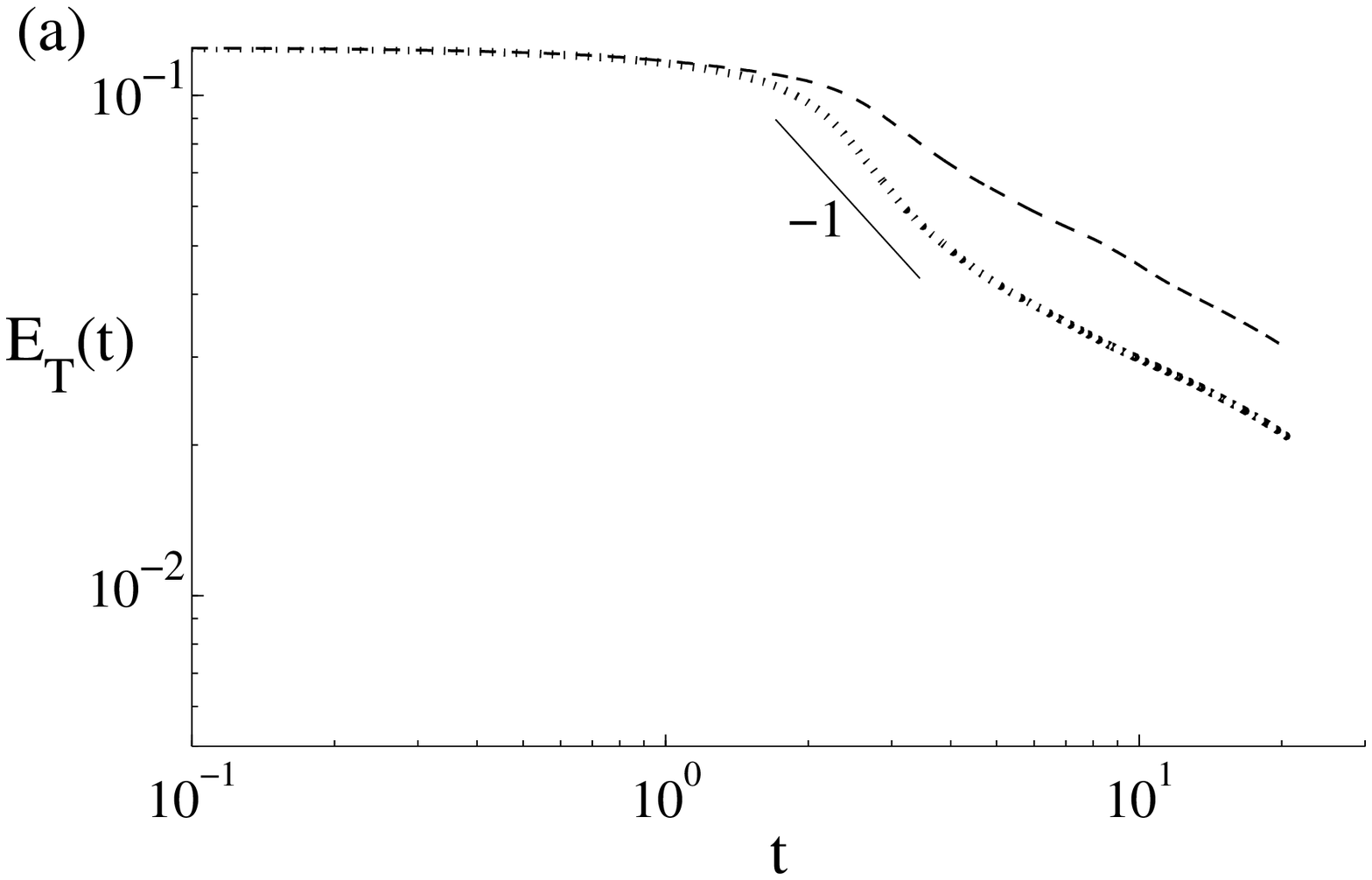}}
\resizebox{7cm}{!}{\includegraphics{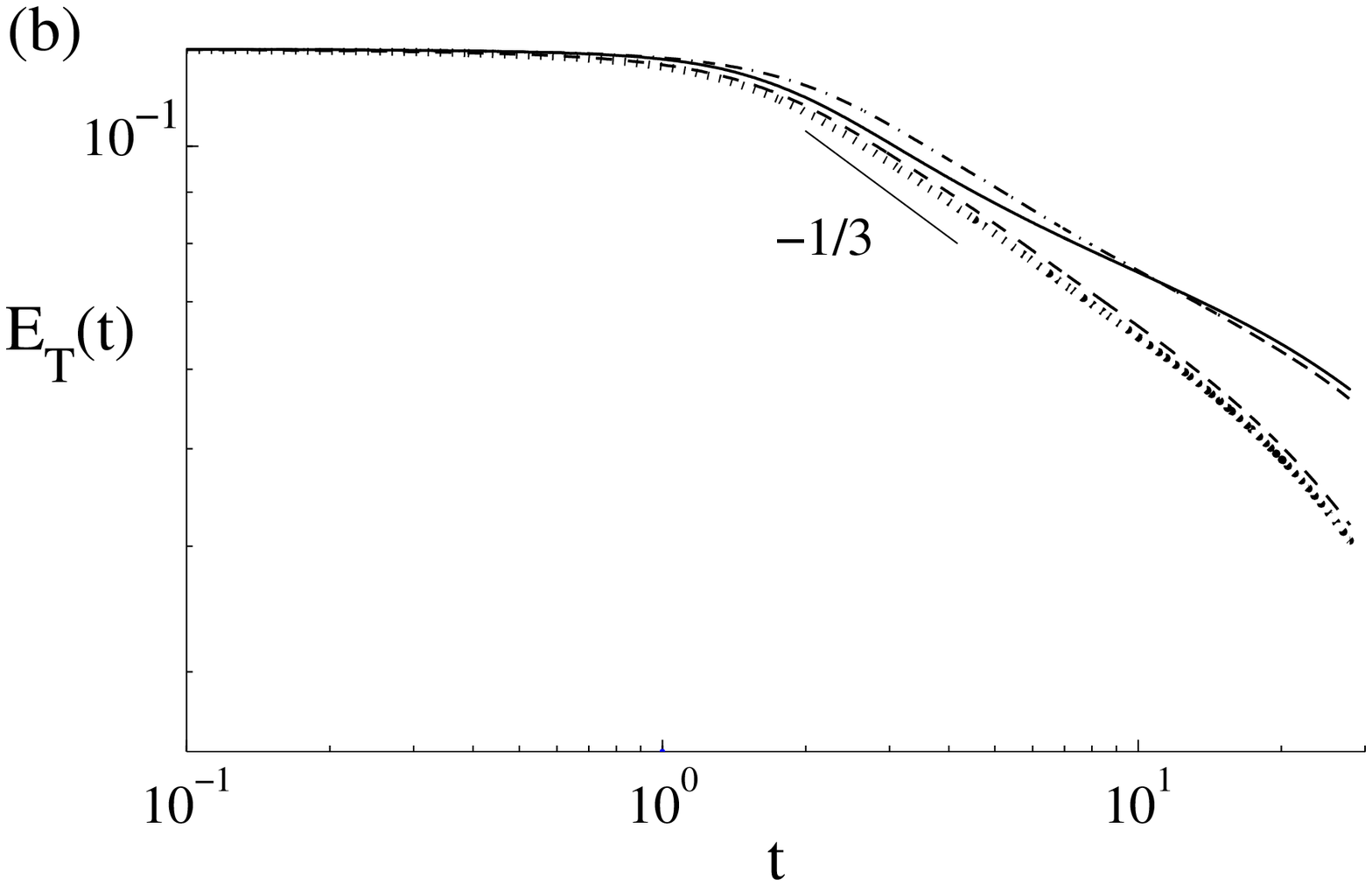}}
\resizebox{7cm}{!}{\includegraphics{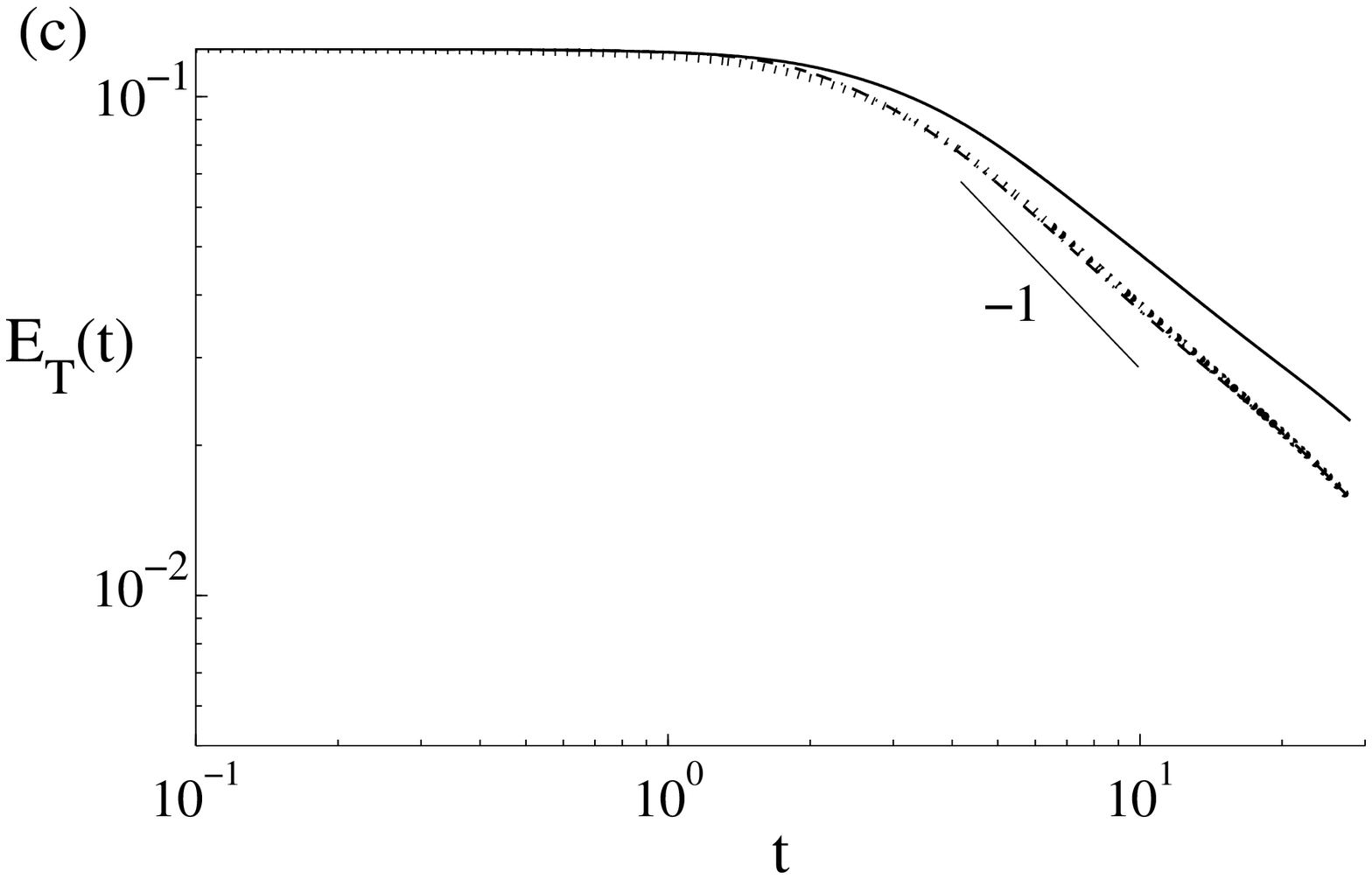}}
\caption{\label{TG_Eta} 
Temporal evolution of the total energy $E_T(t)$, with different scaling laws given as indications for: (a) TG flows at different Froude numbers: $Fr\approx 0.011$ (run 2, TG$Fr/2$, dashed line), and $Fr\approx 0.022$ (run 1, TG, dotted line); (b) ABC flows with $Re=6000$ and $Fr=0.022$ (run 8, ABC$2Re$, dash-dotted line), $Re=6000$ and $Fr=0.044$ (run 9, ABC$2Fr$, solid line), $Re=3000$ and $Fr=0.011$ (run 11, ABC$Fr/2$ dashed line), and $Re=3000$ and $Fr=0.022$ (run 7, ABC, dotted line); (c) random flows with $Re=6000$ and $Fr=0.022$ (run 14, RND$2Re$, solid line), $Re=6000$ and $Fr=0.044$ (run 15, RND$2Fr$, dash-dotted line), and $Re=3000$ and $Fr=0.022$ for reference (run 13, RND, dotted line). Note the significantly slower decay in (b) for the case of the ABC flow.}
 \end{figure*}
 
We now turn our attention to the temporal evolution of energy. We display in Figs.~\ref{TG_Etb}  and \ref{TG_Eta} the decay of the total energy for a variety of initial conditions and Froude numbers (see captions for details). Computations are performed for roughly 30 turnover times (and thus, for $3000$ Brunt-Va\"iss\"al\"a periods, for $Fr=0.01$); some power-laws are added to guide the eye. When examining separately the temporal decay of the potential and kinetic energy, they evolve in similar ways but with strong oscillatory energy exchanges, while the oscillations due to gravity waves disappear when considering the total energy.

In Fig. \ref{TG_Etb}, it is striking to notice that the decay of energy can be very different for different flows with the same external parameters (i.e., Reynolds and Froude numbers). If $t^*$ is the time at which dissipation sets in, that is, the maximum of enstrophy, the decay, in the absence of stratification, would follow a $\sim (t^*-t)^{-2}$ law (dotted line) given that the growth of the integral scale is prevented by being in a ``box-limited'' case. Considering the non-helical TG flow, the decay seems to follow a power-law $\sim (t^*-t)^{-1}$, after an initial ideal (inviscid) phase. This result is expected on the basis of slowing-down of nonlinear interactions (leading to energy transfer and its decay) because of waves, and a similar power-law decay has already been observed for stratified flows \cite{kimura}, for rotating flows \cite{teitelbaum}, as well as for flows in the presence of a magnetic field \cite{galtier_mhd}. 
For the ABC stratified flow the decay of energy is substantially slower, with a power law of $-1/3$ as opposed to $-1$. On the other hand, we recall that helicity decays linearly and at a very slow rate, as shown in the preceding section.

The slow decay of energy has been observed previously for rotating turbulence in the presence of helicity and the different power laws one may expect have been reviewed for a variety of cases taking into account the invariance of both energy and helicity \cite{pablo_thomas}. A $(t^*-t)^{-1/3}$ law is found on phenomenological grounds for helical rotating flows based on the fact that helicity plays a role in the dynamics: it dominates the energy transfer to small scales and alters the spatial scaling laws for the energy spectra and higher order structure functions as well \cite{1536a, 1536b} (see also \cite{trieste} for review). The similar decay law found in this paper for ABC-like stratified flows may be due to the fact that, as shown here, the helicity is quasi-conserved by the dynamics, due to a cyclostrophic balance, and thus leads to the same slow decay. We thus conclude that, for flows with either rotation or stratification, the presence of helicity considerably slows-down the energy decay, and measurably so, leading to persistent structures, whereas for the unstratified non-rotating case, helicity delays the onset of the decay but does not alter its rate of decrease \cite{pablo_thomas}.

In Fig.~\ref{TG_Eta} we examine the possible variation of the decay rate of energy with several factors: Froude and Reynolds numbers, as well as TG, ABC, and random initial conditions. We observe that for the TG flow, the decay is first faster ($\sim t^{-1}$) and then slows down; this could be due to the fact that, as time evolves and energy decays, the Froude number of a given flow becomes substantially smaller; note that the later decay in Fig.~\ref{TG_Eta}(a) corresponds to an exponential regime, as can be easily seen displaying the data in lin-log coordinates (not shown). { If the Froude number is decreased the decay is delayed (dashed-line), while still following the same power law. A similar effect is observed for ABC flows Fig.~\ref{TG_Eta}(b) where results for two different Reynolds numbers are reported: the higher the Reynolds number, the shallower the energy decay.}

For the random cases [Fig.~\ref{TG_Eta}(c)], when $k_{0}=3-4$, helicity has become negligible by the peak of the enstrophy, as previously observed, and the decay follows an approximate $t^{-1}$ law. When $k_{0}=2$  there are too few modes to excite in building a random initial condition, and at early times the randomness of such flows is rather weak, with well-ordered columns in the initial conditions, reminiscent of the ABC flow itself (see, e.g., Fig.~\ref{VAPORvx}). Helicity then does not go to zero, and the decay follows, as in the ABC case, a $-1/3$ power law (this case is not shown). 

In general terms, it seems reasonable to conclude, for stratified flows, that if helicity vanishes by the time of occurrence of the peak of the enstrophy, the decay will follow a -1 power law, while if the initial helicity value is high and it survives up until the maximum of dissipation, the decay will follow a -1/3 power law.

\subsection{Energy spectra} \label{subsec:spectra}
\begin{figure*}[h!tbp] \centering
\resizebox{12cm}{!}{\includegraphics{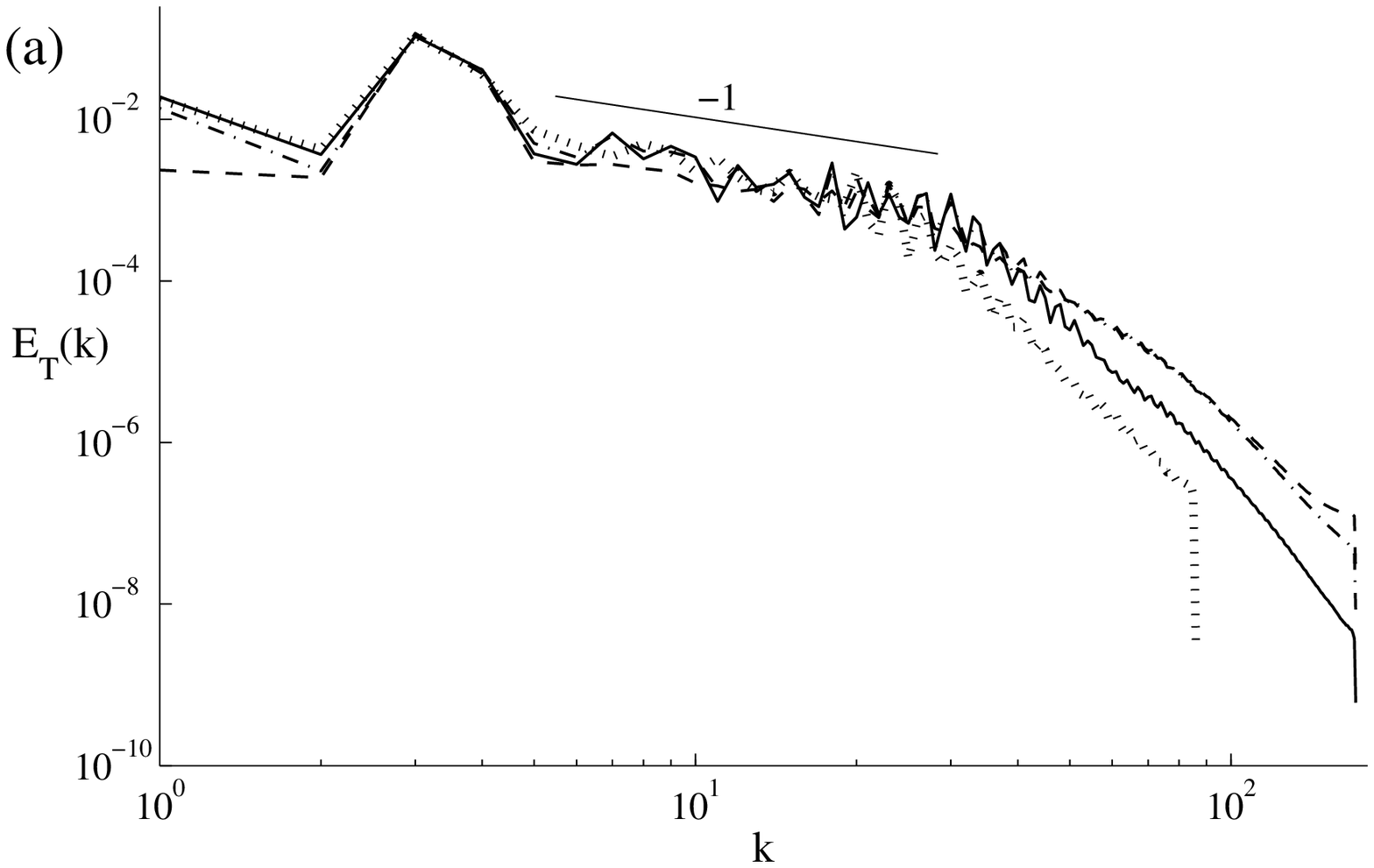}}
\resizebox{12cm}{!}{\includegraphics{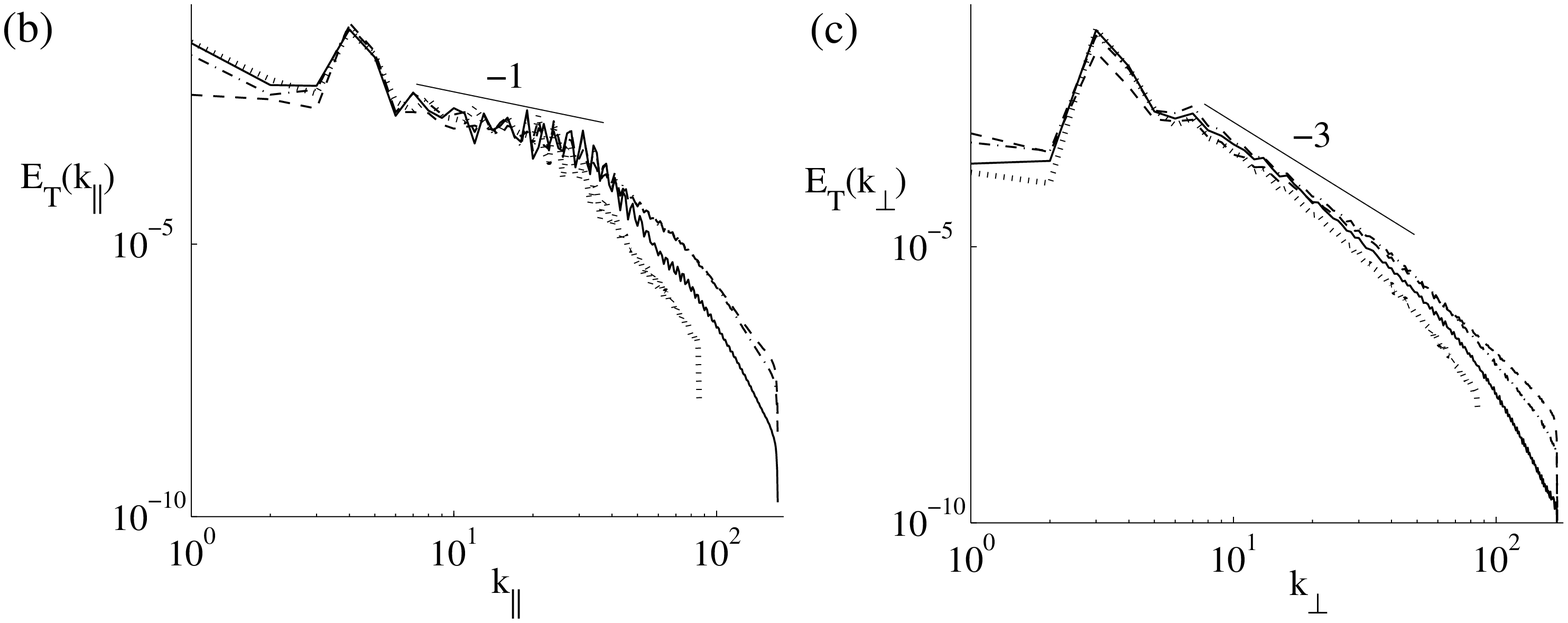}}
\resizebox{12cm}{!}{\includegraphics{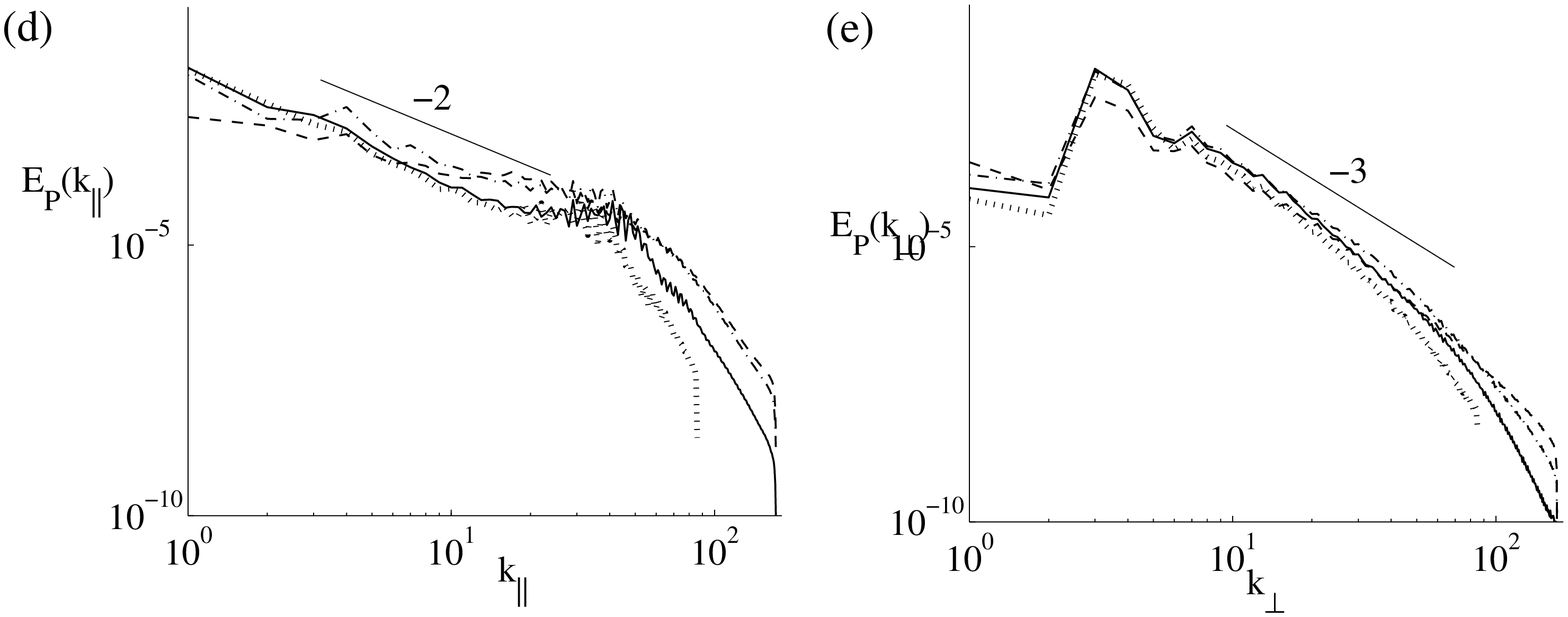}}
\caption{\label{ABCs_spectrum}
Isotropic (a), parallel (b), and perpendicular (c) total energy spectra, and parallel (d) and perpendicular (e) potential energy spectra for ABC flows with $Re=6000$ and $Fr=0.022$ (run 8, ABC$2Re$, solid line), $Re=6000$ and $Fr=0.044$ (run 9, ABC$2Fr$, dash-dotted line), $Re=6000$ and $Fr=0.088$ (run 10, ABC$4Fr$ dashed line), and $Re=3000$ and $Fr=0.022$ (run 7, ABC, dotted line). Slopes are indicated as a reference.
{Note that the rather flat spectrum in the total energy stems from vertical variations (left, b and d), whereas the perpendicular  spectra (right, c and e) are close to a $k_{\parallel}^{-3}$ scaling for all values of parameters.}
} \end{figure*}
\begin{figure*}[h!tbp] \centering
\resizebox{10cm}{!}{\includegraphics{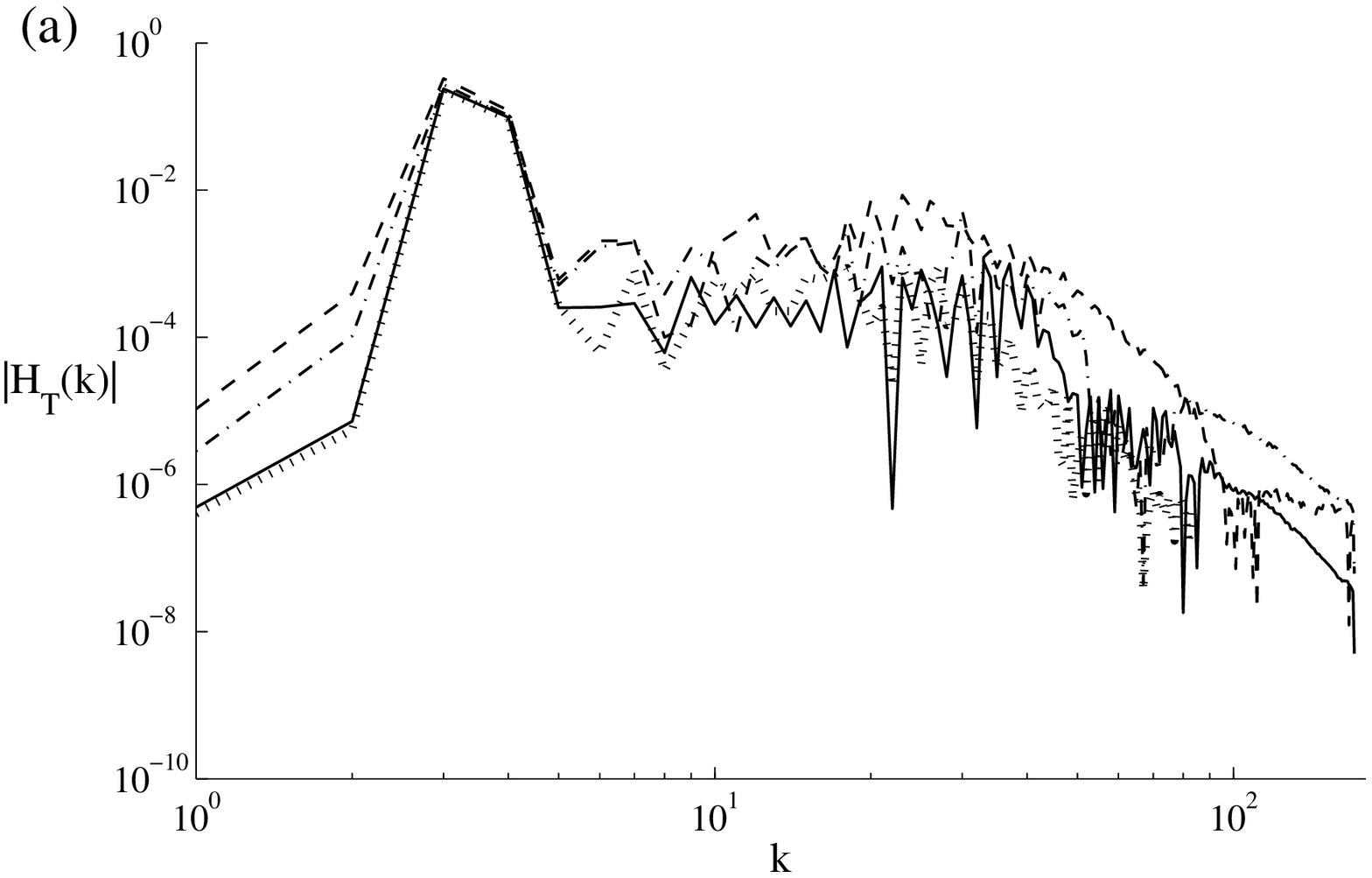}}
\resizebox{12cm}{!}{\includegraphics{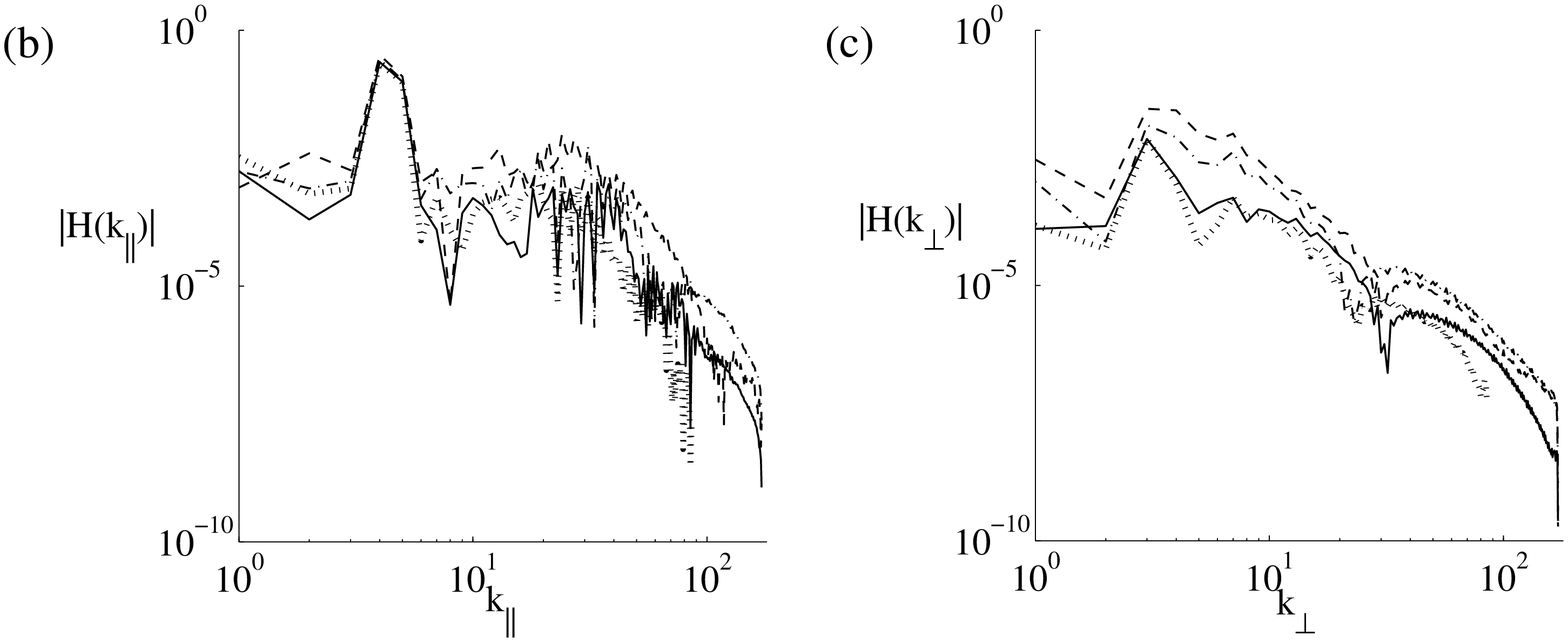}}
\caption{\label{ABCs_Hspectrum}
Isotropic (a), parallel (b), and perpendicular (c) helicity spectra 
{for the same flows as in Fig. \ref{ABCs_spectrum}: run 8 (solid line), run 9 (dash-dotted line), run 10 (dashed line) and run 7 (dotted line). Note the flat spectra at large scales, stemming again from vertical variations.}
 } \end{figure*}
\begin{figure*}[h!tbp] \centering
\resizebox{12cm}{!}{\includegraphics{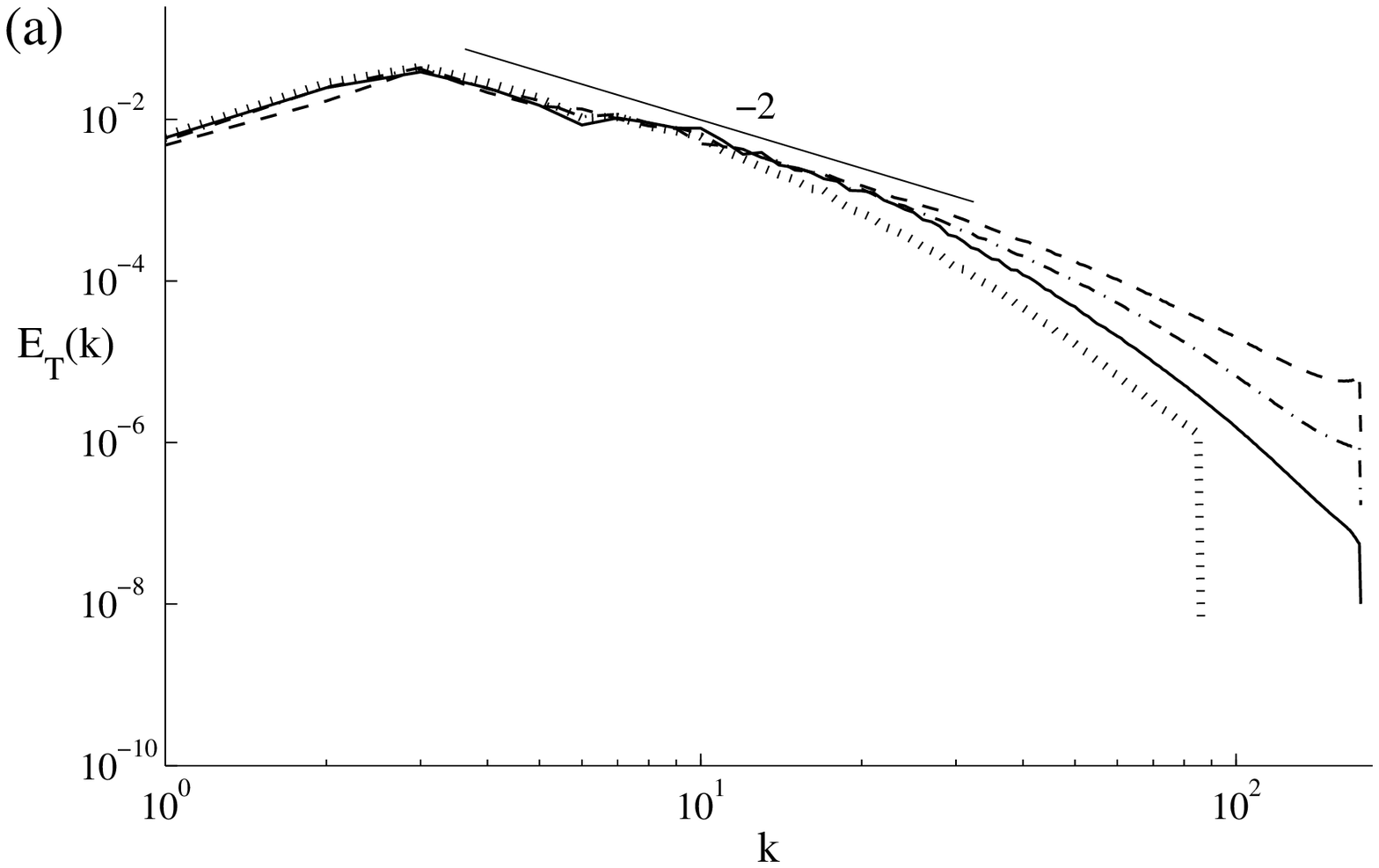}}
\resizebox{12cm}{!}{\includegraphics{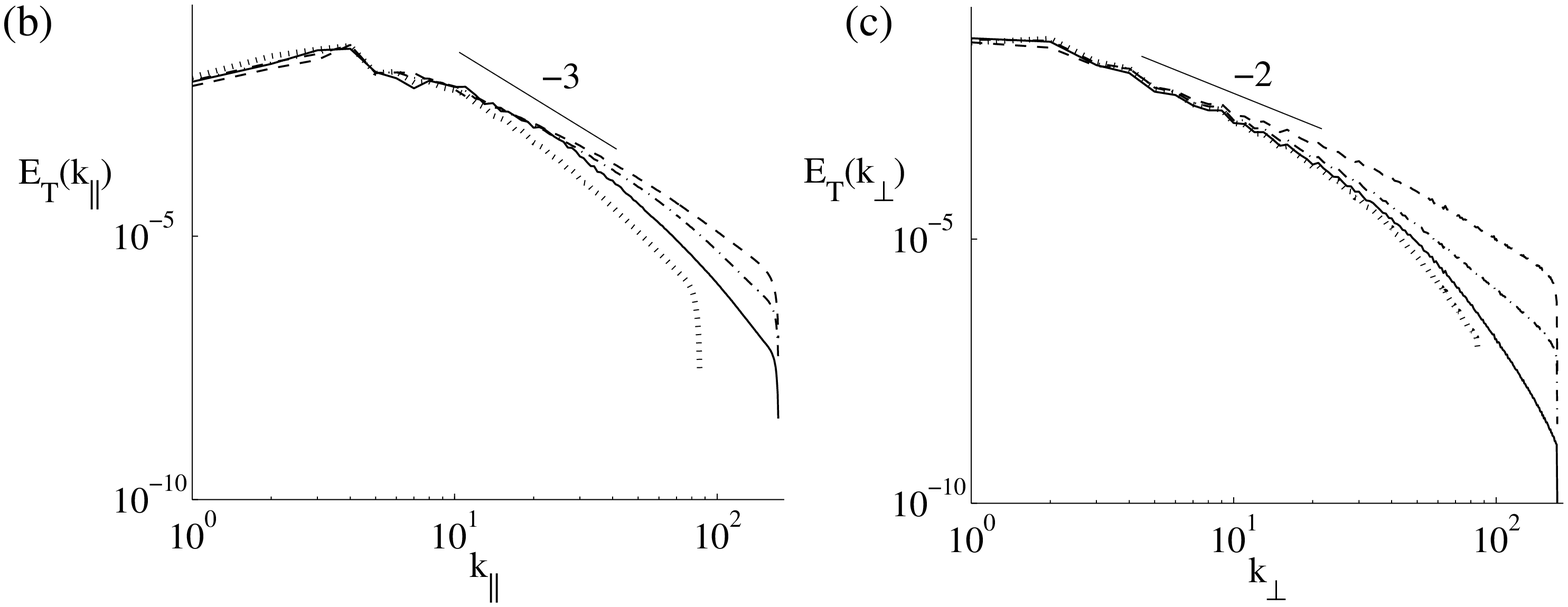}}
\resizebox{12cm}{!}{\includegraphics{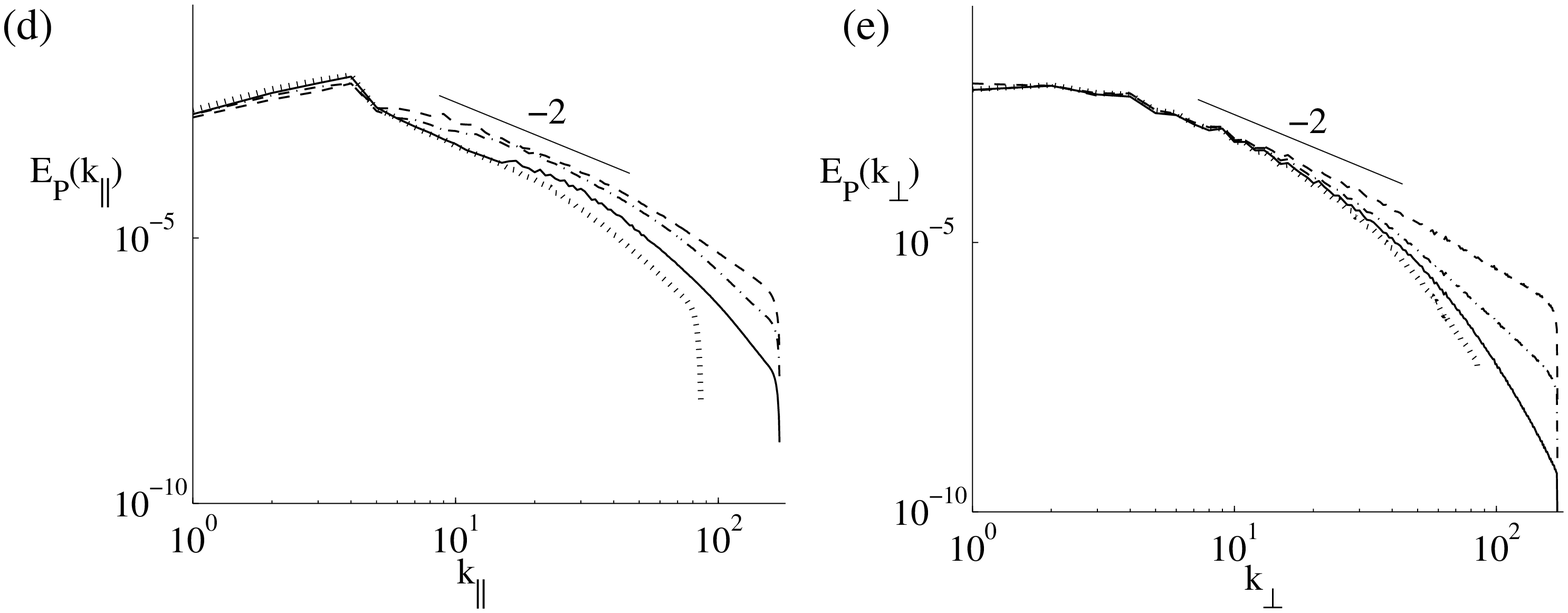}}
\caption{\label{RDMs_spectrum}
Isotropic (a), parallel (b), and perpendicular (c) total energy spectra, and parallel (d) and perpendicular (e) potential energy spectra for random flows with $Re=6000$ and $Fr=0.022$ (run 14, RND$2Re$, solid line), $Re=6000$ and $Fr=0.044$ (run 15, RND$2Fr$, dash-dotted line),  $Re=6000$ and $Fr=0.088$ (run 16, RND$4Fr$, dashed line), and $Re=3000$ and $Fr=0.022$ (run 13, RND, dotted line). Slopes are indicated as a reference.
{No knee in the spectra is discernible, contrary to ABC flows.}
} \end{figure*}
\begin{figure*}[h!tbp] \centering
\resizebox{10cm}{!}{\includegraphics{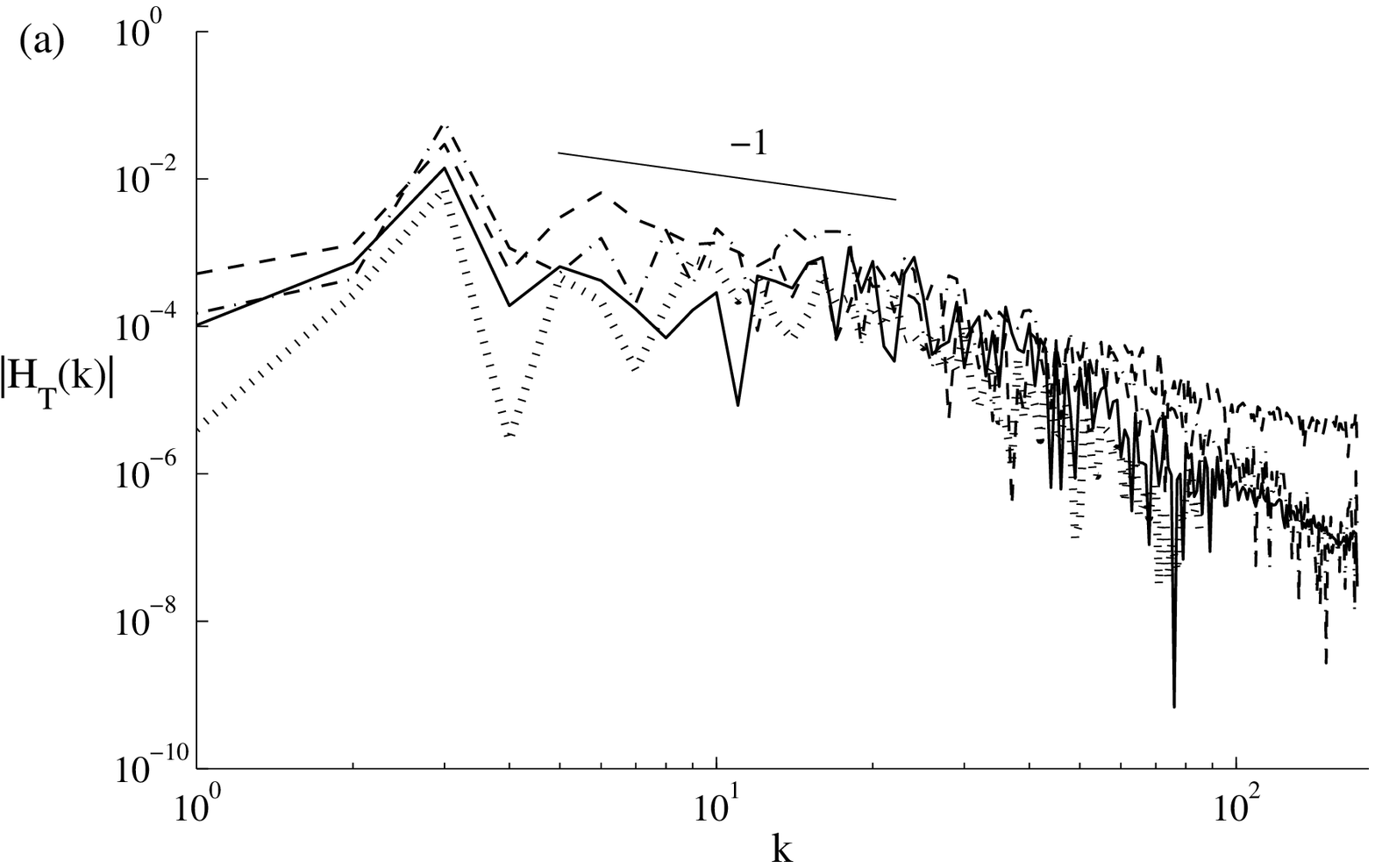}}
\resizebox{12cm}{!}{\includegraphics{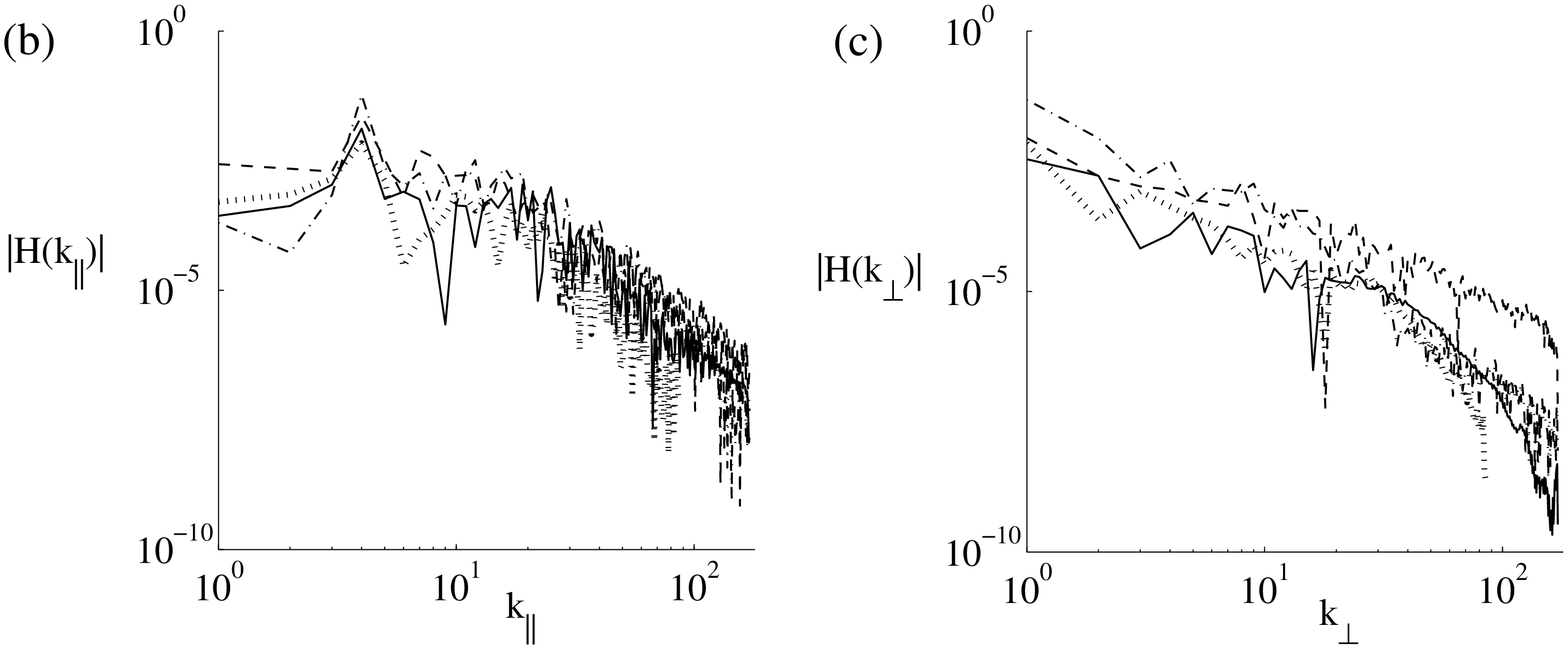}}
\caption{\label{RDMs_Hspectrum}
Isotropic (a), parallel (b), and perpendicular (c) helicity spectra for the same random flows as in Fig. \ref{RDMs_spectrum}: run 14 (solid line), run 15 (dash-dotted line), run 16 (dashed line) and run 13 (dotted line).
 } \end{figure*}

The effect of strong and large scale helicity on the temporal decay rate of the total energy is thus striking. We now address the related question of the possible role of helicity on energy distribution among Fourier modes.

In Fig.~\ref{ABCs_spectrum} we show the isotropic ($k$), parallel ($k_{\parallel}$) and perpendicular ($k_{\perp}$) total energy spectra, $E_T$, together with the parallel and perpendicular potential energy spectra, $E_P$, for four ABC flows with different Froude and Reynolds numbers (see caption). The spectra are plotted at the peak of enstrophy, and are averaged over $\approx 0.3-0.4$ eddy turnover times. 
As general remarks, we note that the energy is rather strongly peaked at $k_{0}$, and we observe that, at high wavenumbers the spectra are more developed when the Froude number is higher, since nonlinear interactions are comparatively more efficient. The survival of the initially excited large scales is expected, since both wave interactions and helicity can inhibit the energy transfer to smaller scales. This fact may explain the rather shallow spectrum at intermediate wavenumbers, followed by the very steep spectrum at high wavenumbers, observed in Fig.~\ref{ABCs_spectrum} (a) and (b).
In the isotropic energy spectrum [Fig~\ref{ABCs_spectrum} (a)], the transition between the rather flat behavior and the steep high-wavenumber scaling law  is marked by a `knee' around $k\approx 30$, which is possibly coincident with the buoyancy wavenumber $k_{b}$: $k_{b}\approx 25$ for $Fr=0.022$, and $k_{b}\approx 13$ for $Fr=0.044$. The same knee is observed in the total and potential parallel energy spectra, being so entirely attributable to the behavior of the flow with respect to variations in the vertical direction, a fact that strengthens its identification with $k_b$.

The perpendicular spectra decay consistently with a $k^{-3}$ power law, as found in several numerical studies for this parameter regime \cite{staquet_rev_02, kimura12}. 

Flat spectra have already been documented in the literature for stratified turbulence and various interpretations have been provided. A flat energy spectrum has been derived in \cite{kimura} 
on the basis of a superposition of independent horizontal layers, each with a Gaussian profile, as a model of the vertically-sheared horizontal layers  associated with the zig-zag instability \cite{chomaz} (as well as with geostrophic balance when one takes rotation into account). Indeed, the thickness of these layers is thought to be the buoyancy scale, as advocated in \cite{chomaz} using a scaling argument that states that the Froude number based on a vertical length scale has to be of order unity. This has already been studied in a series of direct numerical simulations (see, e.g., \cite{godeferd, waite2004} for more details). Here we hypothesize that flat spectra may be related to the presence of high helicity values as we discuss below commenting on Fig.~\ref{ABCs_Hspectrum}. 

{The knee we have identified in the evolution of energy spectra as a function of $k_{\parallel}$ is also not a new feature of stratified turbulence. It has been attributed to a change of regime for the total energy (or its horizontal component) at high resolution (specifically  with the dissipation wave number sufficiently resolved) at the buoyancy scale \cite{waite2011}. On the other hand, in \cite{kimura12}, both the wave mode (corresponding to the vertical velocity) and the vortex mode (corresponding to the horizontal velocity) display a knee in terms of $k_{\parallel}$ for highly stratified flows, identified as the buoyancy scale. Our data leads us to argue (see below) that another, rather independent phenomenon, may be happening as well, namely the role played by the helicity distribution among modes.} 

The isotropic, parallel and perpendicular helicity spectra are shown in  Fig. \ref{ABCs_Hspectrum} for the same ABC flows (note that the absolute value of helicity is displayed). Noteworthy, similarly to the total and parallel isotropic energy spectra are the following points: (i) the spectrum is flat at large scale; and (ii)  a knee is present for the total and parallel helicity; in addition we also observe a secondary change in the spectra at a smaller scale which seems to coincide with a change of sign of helicity. {Since, according to the Schwartz inequality we have $H(k)\leqslant kE(k)$, and since helicity is large for these flows, it is reasonable to think that the behavior of the energy spectra is influenced by the behavior of helicity.} 

As the temporal data indicates that the ABC flow behaves differently from random flows, we now examine the spectral behavior for the random case. In Fig.~\ref{RDMs_spectrum} the energy spectra for random flows centered initially on wavenumber $k_{0}=3,4$ are presented. We recall that at the peak of enstrophy, the residual helicity is close to zero for these runs. In these cases there is no clear signature of a knee or an abrupt slope change for the energy. The higher the $Re$ and  $Fr$ numbers, the shallower the spectra at small scales. The peak at $k_{0}$ is also noticeably reduced with respect to the ABC cases, indicating more efficient energy transfer from the energy-containing scales.

Finally, in Fig.~\ref{RDMs_Hspectrum} the helicity spectra for the random cases initially centered at $k_0={3,4}$ are shown. The value of helicity is much smaller than for the ABC runs, as noted in Table \ref{peak_enstrophy}: the residual relative helicity $\sigma_V$ at the peak of enstrophy is close to 2\%, whereas for the ABC runs it is close to 20\%. At the end of the runs, these values are respectively $\approx11\%$ and $\approx 80\%$, i.e., almost as strong as it can be in the latter case. A  slope change and a knee are still identifiable at a wavenumber again comparable to the buoyancy wavenumber $k_{b}$, but the helicity is now too weak to influence the behavior of the energy spectrum. Note also that at scales larger than the buoyancy scale (but smaller than the initial condition scale), the helicity spectrum is again flat, 
likely due to the layered structure in the vertical, whereas the perpendicular spectrum is steep.

To conclude, it seems likely, for helical flows here represented only by ABC initial conditions, that it is the helicity which is responsible for the change of slope in the energy spectrum close to the buoyancy scale, and not the other way round. Assuming that the Schwarz inequality is fulfilled, a flat helicity spectrum leads to $E(k)\sim k^{-1}$, compatible with what we observe for the energy in the ABC case. A similar result obtains for the random initial condition centered on $k_0={2}$. This point will need further studies at high resolution so that different energy and helicity balance can take place in flows with sufficient scale separation.

\section{Conclusion} \label{sec:conclusion}

We have performed numerical simulations of stratified turbulence and have shown that initial conditions with sufficiently strong large-scale helicity produce measurable effects. This is observable in both the energy decay, which is found to be substantially slower than in the non-helical case, and in the energy (and helicity) spectra, which display a flat distribution with a marked knee in the vicinity of the buoyancy scale $L_b$; beyond $L_b$, the spectra are significantly steeper for the cases examined in this paper, which are strongly stratified, with buoyancy Reynolds number of order unity and up to 46 (see Table \ref{simulations}).
The knees are clearly identified with variations of these spectra with the vertical wavenumber, and the knees  seem to be close to the buoyancy scale, in agreement with the interpretation of $L_b$ as the scale at which $Fr\sim 1$.

{Thus, for helical stratified turbulence, the energy decay is slowed down by the presence of helicity in a way reminiscent to that observed in the rotating case. {However, it should be noted that helicity is an ideal invariant of rotating flows, whereas it is not {\it a priori} conserved for ideal stratified flows. While in the former case the slow down in the decay is associated with a direct cascade of the helicity, in the latter case a very slow (linear in time) decay of helicity (resulting in a quasi-conservation)} can be attributed to a cyclostrophic balance where dissipation and gravity equilibrate at large scales once the nonlinear terms start transferring energy to small scales after an initial wave-dominated regime. These results are obtained in cases where the evolution of large scales is somewhat inhibited, because  the initial scale of the flow is comparable to the overall size of the computational box, the so-called box-limited conditions in which the growth of the integral scale is blocked.
}

{Moreover, in highly turbulent flows such as those found in geophysics and astrophysics, the eddy turn-over time becomes shorter than $\tau_W$  at smaller scales, for a sufficiently large Reynolds number. In that case,  isotropic turbulence  may recover, beyond the so-called Ozmidov scale 
$L_{oz}\sim [\epsilon/N^3]^{1/2}$.  
Assuming further that this scale is larger than the dissipation scale $\ell_{diss}\sim [\epsilon/\nu^3]^{1/4}$ leads to the condition that the buoyancy Reynolds number ${\cal R}=ReFr^2$ be larger than unity.
It was found recently that power laws of the energy spectra scale with  ${\cal R}$, as opposed to the Froude number itself \cite{bartello_13}.
The condition ${\cal R}>> 1$ is easily fulfilled in geophysical flows but difficult to realize in numerical simulations, where the highest to date is found in \cite{almalkie_12}, realized on a grid of $4096^2*1024$ points; similarly, laboratory experiments function at moderate ${\cal R}$  \cite{waite_2012}. Further studies of that regime are necessary but will require substantial numerical resources. }
\acknowledgements
This work is  supported by NSF/CMG grant 1025183.
This work was also sponsored by an NSF cooperative agreement through the University Corporation for Atmospheric Research on behalf of the National Center for Atmospheric Research (NCAR). Computer time was provided by NSF under sponsorship of NCAR. Additional computational resources were provided by NSF-MRI Grant CNS-0821794, MRI-Consortium: Acquisition of a Supercomputer by the Front Range Computing 
Consortium (FRCC), with additional support from the University of Colorado and NSF sponsorship of NCAR. Cecilia Rorai was supported by a graduate research grant from the Advanced Study Program at NCAR, and from a RSVP/CISL grant at NCAR. 
\vskip0.3truein

\bibliography{arxiv_rorai12.bib}

\begin{thebibliography}{55}%
\makeatletter
\providecommand \@ifxundefined [1]{%
 \@ifx{#1\undefined}
}%
\providecommand \@ifnum [1]{%
 \ifnum #1\expandafter \@firstoftwo
 \else \expandafter \@secondoftwo
 \fi
}%
\providecommand \@ifx [1]{%
 \ifx #1\expandafter \@firstoftwo
 \else \expandafter \@secondoftwo
 \fi
}%
\providecommand \natexlab [1]{#1}%
\providecommand \enquote  [1]{``#1''}%
\providecommand \bibnamefont  [1]{#1}%
\providecommand \bibfnamefont [1]{#1}%
\providecommand \citenamefont [1]{#1}%
\providecommand \href@noop [0]{\@secondoftwo}%
\providecommand \href [0]{\begingroup \@sanitize@url \@href}%
\providecommand \@href[1]{\@@startlink{#1}\@@href}%
\providecommand \@@href[1]{\endgroup#1\@@endlink}%
\providecommand \@sanitize@url [0]{\catcode `\\12\catcode `\$12\catcode
  `\&12\catcode `\#12\catcode `\^12\catcode `\_12\catcode `\%12\relax}%
\providecommand \@@startlink[1]{}%
\providecommand \@@endlink[0]{}%
\providecommand \url  [0]{\begingroup\@sanitize@url \@url }%
\providecommand \@url [1]{\endgroup\@href {#1}{\urlprefix }}%
\providecommand \urlprefix  [0]{URL }%
\providecommand \Eprint [0]{\href }%
\providecommand \doibase [0]{http://dx.doi.org/}%
\providecommand \selectlanguage [0]{\@gobble}%
\providecommand \bibinfo  [0]{\@secondoftwo}%
\providecommand \bibfield  [0]{\@secondoftwo}%
\providecommand \translation [1]{[#1]}%
\providecommand \BibitemOpen [0]{}%
\providecommand \bibitemStop [0]{}%
\providecommand \bibitemNoStop [0]{.\EOS\space}%
\providecommand \EOS [0]{\spacefactor3000\relax}%
\providecommand \BibitemShut  [1]{\csname bibitem#1\endcsname}%
\let\auto@bib@innerbib\@empty
\bibitem [{\citenamefont {Riley}\ and\ \citenamefont
  {Lelong}(2000)}]{riley_rev_00}%
  \BibitemOpen
  \bibfield  {author} {\bibinfo {author} {\bibfnamefont {J.}~\bibnamefont
  {Riley}}\ and\ \bibinfo {author} {\bibfnamefont {M.-P.}\ \bibnamefont
  {Lelong}},\ }\href@noop {} {\bibfield  {journal} {\bibinfo  {journal} {Ann.
  Rev. Fluid Mech.}\ }\textbf {\bibinfo {volume} {{\bf 32}}},\ \bibinfo {pages}
  {249} (\bibinfo {year} {2000})}\BibitemShut {NoStop}%
\bibitem [{\citenamefont {Staquet}\ and\ \citenamefont
  {Sommeria}(2002)}]{staquet_rev_02}%
  \BibitemOpen
  \bibfield  {author} {\bibinfo {author} {\bibfnamefont {C.}~\bibnamefont
  {Staquet}}\ and\ \bibinfo {author} {\bibfnamefont {J.}~\bibnamefont
  {Sommeria}},\ }\href@noop {} {\bibfield  {journal} {\bibinfo  {journal} {Ann.
  Rev. Fluid Mech.}\ }\textbf {\bibinfo {volume} {{\bf 34}}},\ \bibinfo {pages}
  {559} (\bibinfo {year} {2002})}\BibitemShut {NoStop}%
\bibitem [{\citenamefont {Liechtenstein}\ \emph {et~al.}(2005)\citenamefont
  {Liechtenstein}, \citenamefont {Godeferd},\ and\ \citenamefont
  {Cambon}}]{cambon_05}%
  \BibitemOpen
  \bibfield  {author} {\bibinfo {author} {\bibfnamefont {L.}~\bibnamefont
  {Liechtenstein}}, \bibinfo {author} {\bibfnamefont {F.}~\bibnamefont
  {Godeferd}}, \ and\ \bibinfo {author} {\bibfnamefont {C.}~\bibnamefont
  {Cambon}},\ }\href@noop {} {\bibfield  {journal} {\bibinfo  {journal} {J. of
  Turbulence}\ }\textbf {\bibinfo {volume} {{\bf 6}}},\ \bibinfo {pages} {1}
  (\bibinfo {year} {2005})}\BibitemShut {NoStop}%
\bibitem [{\citenamefont {Peltier}\ and\ \citenamefont
  {Caulfield}(2003)}]{peltier_03}%
  \BibitemOpen
  \bibfield  {author} {\bibinfo {author} {\bibfnamefont {W.}~\bibnamefont
  {Peltier}}\ and\ \bibinfo {author} {\bibfnamefont {C.}~\bibnamefont
  {Caulfield}},\ }\href@noop {} {\bibfield  {journal} {\bibinfo  {journal}
  {Ann. Rev. Fluid Mech.}\ }\textbf {\bibinfo {volume} {{\bf 35}}},\ \bibinfo
  {pages} {135} (\bibinfo {year} {2003})}\BibitemShut {NoStop}%
\bibitem [{\citenamefont {Ivey}\ \emph {et~al.}(2008)\citenamefont {Ivey},
  \citenamefont {Winters},\ and\ \citenamefont {Koseff}}]{ivey_08}%
  \BibitemOpen
  \bibfield  {author} {\bibinfo {author} {\bibfnamefont {G.}~\bibnamefont
  {Ivey}}, \bibinfo {author} {\bibfnamefont {K.}~\bibnamefont {Winters}}, \
  and\ \bibinfo {author} {\bibfnamefont {J.}~\bibnamefont {Koseff}},\
  }\href@noop {} {\bibfield  {journal} {\bibinfo  {journal} {Ann. Rev. Fluid
  Mech.}\ }\textbf {\bibinfo {volume} {{\bf 40}}},\ \bibinfo {pages} {169}
  (\bibinfo {year} {2008})}\BibitemShut {NoStop}%
\bibitem [{\citenamefont {Embid}\ and\ \citenamefont
  {Majda}(1998)}]{embid_majda_98}%
  \BibitemOpen
  \bibfield  {author} {\bibinfo {author} {\bibfnamefont {P.}~\bibnamefont
  {Embid}}\ and\ \bibinfo {author} {\bibfnamefont {A.}~\bibnamefont {Majda}},\
  }\href@noop {} {\bibfield  {journal} {\bibinfo  {journal} {Geophys.
  Astrophys. Fluid Dyn.}\ }\textbf {\bibinfo {volume} {{\bf 87}}},\ \bibinfo
  {pages} {1} (\bibinfo {year} {1998})}\BibitemShut {NoStop}%
\bibitem [{\citenamefont {Julien}\ \emph {et~al.}(2006)\citenamefont {Julien},
  \citenamefont {Knobloch}, \citenamefont {Milliff},\ and\ \citenamefont
  {Werne}}]{julien_06}%
  \BibitemOpen
  \bibfield  {author} {\bibinfo {author} {\bibfnamefont {K.}~\bibnamefont
  {Julien}}, \bibinfo {author} {\bibfnamefont {E.}~\bibnamefont {Knobloch}},
  \bibinfo {author} {\bibfnamefont {R.}~\bibnamefont {Milliff}}, \ and\
  \bibinfo {author} {\bibfnamefont {J.}~\bibnamefont {Werne}},\ }\href@noop {}
  {\bibfield  {journal} {\bibinfo  {journal} {J. Fluid Mech.}\ }\textbf
  {\bibinfo {volume} {{\bf 555}}},\ \bibinfo {pages} {233} (\bibinfo {year}
  {2006})}\BibitemShut {NoStop}%
\bibitem [{\citenamefont {Majda}\ \emph {et~al.}(2008)\citenamefont {Majda},
  \citenamefont {Mohammadian},\ and\ \citenamefont {Xing}}]{majda_08}%
  \BibitemOpen
  \bibfield  {author} {\bibinfo {author} {\bibfnamefont {A.~J.}\ \bibnamefont
  {Majda}}, \bibinfo {author} {\bibfnamefont {M.}~\bibnamefont {Mohammadian}},
  \ and\ \bibinfo {author} {\bibfnamefont {Y.}~\bibnamefont {Xing}},\
  }\href@noop {} {\bibfield  {journal} {\bibinfo  {journal} {GAFD}\ }\textbf
  {\bibinfo {volume} {{\bf 102}}},\ \bibinfo {pages} {543} (\bibinfo {year}
  {2008})}\BibitemShut {NoStop}%
\bibitem [{\citenamefont {Klein}(2010)}]{klein_rev_10}%
  \BibitemOpen
  \bibfield  {author} {\bibinfo {author} {\bibfnamefont {R.}~\bibnamefont
  {Klein}},\ }\href@noop {} {\bibfield  {journal} {\bibinfo  {journal} {Ann.
  Rev. Fluid Mech.}\ }\textbf {\bibinfo {volume} {{\bf 42}}},\ \bibinfo {pages}
  {613} (\bibinfo {year} {2010})}\BibitemShut {NoStop}%
\bibitem [{\citenamefont {Grooms}\ \emph {et~al.}(2010)\citenamefont {Grooms},
  \citenamefont {Julien},\ and\ \citenamefont {Knobloch}}]{grooms_10}%
  \BibitemOpen
  \bibfield  {author} {\bibinfo {author} {\bibfnamefont {I.}~\bibnamefont
  {Grooms}}, \bibinfo {author} {\bibfnamefont {K.}~\bibnamefont {Julien}}, \
  and\ \bibinfo {author} {\bibfnamefont {E.}~\bibnamefont {Knobloch}},\
  }\href@noop {} {\bibfield  {journal} {\bibinfo  {journal} {PRL}\ }\textbf
  {\bibinfo {volume} {{\bf 104}}},\ \bibinfo {pages} {224501} (\bibinfo {year}
  {2010})}\BibitemShut {NoStop}%
\bibitem [{\citenamefont {Wingate}\ \emph {et~al.}(2011)\citenamefont
  {Wingate}, \citenamefont {Embid}, \citenamefont {Holmes-Cerfon},\ and\
  \citenamefont {Taylor}}]{wingate_11}%
  \BibitemOpen
  \bibfield  {author} {\bibinfo {author} {\bibfnamefont {B.~A.}\ \bibnamefont
  {Wingate}}, \bibinfo {author} {\bibfnamefont {P.}~\bibnamefont {Embid}},
  \bibinfo {author} {\bibfnamefont {M.}~\bibnamefont {Holmes-Cerfon}}, \ and\
  \bibinfo {author} {\bibfnamefont {M.~A.}\ \bibnamefont {Taylor}},\
  }\href@noop {} {\bibfield  {journal} {\bibinfo  {journal} {J. Fluid Mech.}\
  }\textbf {\bibinfo {volume} {{\bf 676}}},\ \bibinfo {pages} {546} (\bibinfo
  {year} {2011})}\BibitemShut {NoStop}%
\bibitem [{\citenamefont {Julien}\ \emph {et~al.}(2012)\citenamefont {Julien},
  \citenamefont {Rubio}, \citenamefont {Grooms},\ and\ \citenamefont
  {Knobloch}}]{julien_12}%
  \BibitemOpen
  \bibfield  {author} {\bibinfo {author} {\bibfnamefont {K.}~\bibnamefont
  {Julien}}, \bibinfo {author} {\bibfnamefont {A.~M.}\ \bibnamefont {Rubio}},
  \bibinfo {author} {\bibfnamefont {I.}~\bibnamefont {Grooms}}, \ and\ \bibinfo
  {author} {\bibfnamefont {E.}~\bibnamefont {Knobloch}},\ }\href@noop {}
  {\bibfield  {journal} {\bibinfo  {journal} {GAFD}\ }\textbf {\bibinfo
  {volume} {in press}} (\bibinfo {year} {2012})}\BibitemShut {NoStop}%
\bibitem [{\citenamefont {Godeferd}\ and\ \citenamefont
  {Cambon}(1994)}]{cambon_94}%
  \BibitemOpen
  \bibfield  {author} {\bibinfo {author} {\bibfnamefont {F.}~\bibnamefont
  {Godeferd}}\ and\ \bibinfo {author} {\bibfnamefont {C.}~\bibnamefont
  {Cambon}},\ }\href@noop {} {\bibfield  {journal} {\bibinfo  {journal} {Phys.
  Fluids}\ }\textbf {\bibinfo {volume} {{\bf 6}}},\ \bibinfo {pages} {2084}
  (\bibinfo {year} {1994})}\BibitemShut {NoStop}%
\bibitem [{\citenamefont {Staquet}\ and\ \citenamefont
  {Godeferd}(1998)}]{staquet_98}%
  \BibitemOpen
  \bibfield  {author} {\bibinfo {author} {\bibfnamefont {C.}~\bibnamefont
  {Staquet}}\ and\ \bibinfo {author} {\bibfnamefont {F.}~\bibnamefont
  {Godeferd}},\ }\href@noop {} {\bibfield  {journal} {\bibinfo  {journal} {J.
  Fluid Mech.}\ }\textbf {\bibinfo {volume} {{\bf 360}}},\ \bibinfo {pages}
  {295} (\bibinfo {year} {1998})}\BibitemShut {NoStop}%
\bibitem [{\citenamefont {Godeferd}\ and\ \citenamefont
  {Staquet}(2003{\natexlab{a}})}]{godef_03}%
  \BibitemOpen
  \bibfield  {author} {\bibinfo {author} {\bibfnamefont {F.}~\bibnamefont
  {Godeferd}}\ and\ \bibinfo {author} {\bibfnamefont {C.}~\bibnamefont
  {Staquet}},\ }\href@noop {} {\bibfield  {journal} {\bibinfo  {journal} {J.
  Fluid Mech.}\ }\textbf {\bibinfo {volume} {{\bf 486}}},\ \bibinfo {pages}
  {115} (\bibinfo {year} {2003}{\natexlab{a}})}\BibitemShut {NoStop}%
\bibitem [{\citenamefont {Nazarenko}(2011)}]{nazar}%
  \BibitemOpen
  \bibfield  {author} {\bibinfo {author} {\bibfnamefont {S.}~\bibnamefont
  {Nazarenko}},\ }\href@noop {} { {\bibinfo {title} {Wave Turbulence}}},\
  Vol.~\bibinfo {volume} {{\bf 825}}\ (\bibinfo  {publisher}
  {Springer-Verlag},\ \bibinfo {year} {2011})\BibitemShut {NoStop}%
\bibitem [{\citenamefont {Newell}\ and\ \citenamefont {Rumpf}(2011)}]{newell}%
  \BibitemOpen
  \bibfield  {author} {\bibinfo {author} {\bibfnamefont {A.}~\bibnamefont
  {Newell}}\ and\ \bibinfo {author} {\bibfnamefont {B.}~\bibnamefont {Rumpf}},\
  }\href@noop {} {\bibfield  {journal} {\bibinfo  {journal} {Ann. Rev. Fluid
  Mech.}\ }\textbf {\bibinfo {volume} {59}},\ \bibinfo {pages} {59} (\bibinfo
  {year} {2011})}\BibitemShut {NoStop}%
\bibitem [{\citenamefont {Caillol}\ and\ \citenamefont
  {Zeitlin}(2000)}]{caillol}%
  \BibitemOpen
  \bibfield  {author} {\bibinfo {author} {\bibfnamefont {P.}~\bibnamefont
  {Caillol}}\ and\ \bibinfo {author} {\bibfnamefont {V.}~\bibnamefont
  {Zeitlin}},\ }\href@noop {} {\bibfield  {journal} {\bibinfo  {journal} {Dyn.
  Atm. Ocean}\ }\textbf {\bibinfo {volume} {{\bf 32}}},\ \bibinfo {pages} {81}
  (\bibinfo {year} {2000})}\BibitemShut {NoStop}%
\bibitem [{\citenamefont {Bretherton}(1964)}]{bretherton}%
  \BibitemOpen
  \bibfield  {author} {\bibinfo {author} {\bibfnamefont {F.}~\bibnamefont
  {Bretherton}},\ }\href@noop {} {\bibfield  {journal} {\bibinfo  {journal} {J.
  Fluid Mech.}\ }\textbf {\bibinfo {volume} {{\bf 20}}},\ \bibinfo {pages}
  {457} (\bibinfo {year} {1964})}\BibitemShut {NoStop}%
\bibitem [{\citenamefont {McComas}\ and\ \citenamefont
  {Bretherton}(1977)}]{mccomas}%
  \BibitemOpen
  \bibfield  {author} {\bibinfo {author} {\bibfnamefont {C.}~\bibnamefont
  {McComas}}\ and\ \bibinfo {author} {\bibfnamefont {F.}~\bibnamefont
  {Bretherton}},\ }\href@noop {} {\bibfield  {journal} {\bibinfo  {journal} {J.
  Geophys. Res.}\ }\textbf {\bibinfo {volume} {{\bf 82}}},\ \bibinfo {pages}
  {1397} (\bibinfo {year} {1977})}\BibitemShut {NoStop}%
\bibitem [{\citenamefont {Majda}\ and\ \citenamefont {Tabak}(1996)}]{majda_96}%
  \BibitemOpen
  \bibfield  {author} {\bibinfo {author} {\bibfnamefont {A.}~\bibnamefont
  {Majda}}\ and\ \bibinfo {author} {\bibfnamefont {E.}~\bibnamefont {Tabak}},\
  }\href@noop {} {\bibfield  {journal} {\bibinfo  {journal} {Physica D}\
  }\textbf {\bibinfo {volume} {{\bf 96}}},\ \bibinfo {pages} {515} (\bibinfo
  {year} {1996})}\BibitemShut {NoStop}%
\bibitem [{\citenamefont {Molemaker}\ \emph {et~al.}(2010)\citenamefont
  {Molemaker}, \citenamefont {McWilliams},\ and\ \citenamefont
  {Capet}}]{mcwilliams_10}%
  \BibitemOpen
  \bibfield  {author} {\bibinfo {author} {\bibfnamefont {M.}~\bibnamefont
  {Molemaker}}, \bibinfo {author} {\bibfnamefont {J.}~\bibnamefont
  {McWilliams}}, \ and\ \bibinfo {author} {\bibfnamefont {X.}~\bibnamefont
  {Capet}},\ }\href@noop {} {\bibfield  {journal} {\bibinfo  {journal} {J.
  Fluid Mech.}\ }\textbf {\bibinfo {volume} {{\bf 654}}},\ \bibinfo {pages}
  {35} (\bibinfo {year} {2010})}\BibitemShut {NoStop}%
\bibitem [{\citenamefont {Kimura}\ and\ \citenamefont
  {Herring}(1996)}]{kimura}%
  \BibitemOpen
  \bibfield  {author} {\bibinfo {author} {\bibfnamefont {Y.}~\bibnamefont
  {Kimura}}\ and\ \bibinfo {author} {\bibfnamefont {J.}~\bibnamefont
  {Herring}},\ }\href@noop {} {\bibfield  {journal} {\bibinfo  {journal} {J.
  Fluid Mech.}\ }\textbf {\bibinfo {volume} {{\bf 328}}},\ \bibinfo {pages}
  {253} (\bibinfo {year} {1996})}\BibitemShut {NoStop}%
\bibitem [{\citenamefont {Almalkie}\ and\ \citenamefont
  {de~Bruyn~Kops}(2012)}]{almalkie_12}%
  \BibitemOpen
  \bibfield  {author} {\bibinfo {author} {\bibfnamefont {S.}~\bibnamefont
  {Almalkie}}\ and\ \bibinfo {author} {\bibfnamefont {S.~M.}\ \bibnamefont
  {de~Bruyn~Kops}},\ }\href@noop {} {\bibfield  {journal} {\bibinfo  {journal}
  {J. Turb}\ }\textbf {\bibinfo {volume} {13}} (\bibinfo {year}
  {2012})}\BibitemShut {NoStop}%
\bibitem [{\citenamefont {Wunsch}\ and\ \citenamefont
  {Ferrari}(2004)}]{wunsch_rev}%
  \BibitemOpen
  \bibfield  {author} {\bibinfo {author} {\bibfnamefont {C.}~\bibnamefont
  {Wunsch}}\ and\ \bibinfo {author} {\bibfnamefont {R.}~\bibnamefont
  {Ferrari}},\ }\href@noop {} {\bibfield  {journal} {\bibinfo  {journal} {Ann.
  Rev. Fluid Mech.}\ }\textbf {\bibinfo {volume} {{\bf 36}}},\ \bibinfo {pages}
  {281} (\bibinfo {year} {2004})}\BibitemShut {NoStop}%
\bibitem [{\citenamefont {MacCready}\ and\ \citenamefont
  {Geyer}(2010)}]{maccready}%
  \BibitemOpen
  \bibfield  {author} {\bibinfo {author} {\bibfnamefont {P.}~\bibnamefont
  {MacCready}}\ and\ \bibinfo {author} {\bibfnamefont {W.~R.}\ \bibnamefont
  {Geyer}},\ }\href@noop {} {\bibfield  {journal} {\bibinfo  {journal} {Ann.
  Rev. Marine Sci.}\ }\textbf {\bibinfo {volume} {2}},\ \bibinfo {pages} {35}
  (\bibinfo {year} {2010})}\BibitemShut {NoStop}%
\bibitem [{\citenamefont {Brandenburg}\ and\ \citenamefont
  {Subramanian}(2005)}]{branden_rev}%
  \BibitemOpen
  \bibfield  {author} {\bibinfo {author} {\bibfnamefont {A.}~\bibnamefont
  {Brandenburg}}\ and\ \bibinfo {author} {\bibfnamefont {K.}~\bibnamefont
  {Subramanian}},\ }\href@noop {} {\bibfield  {journal} {\bibinfo  {journal}
  {Phys. Rep.}\ }\textbf {\bibinfo {volume} {{\bf 417}}} (\bibinfo {year}
  {2005})}\BibitemShut {NoStop}%
\bibitem [{\citenamefont {Rassmussen}\ and\ \citenamefont
  {Blanchard}(1998)}]{moli}%
  \BibitemOpen
  \bibfield  {author} {\bibinfo {author} {\bibfnamefont {E.}~\bibnamefont
  {Rassmussen}}\ and\ \bibinfo {author} {\bibfnamefont {D.}~\bibnamefont
  {Blanchard}},\ }\href@noop {} {\bibfield  {journal} {\bibinfo  {journal}
  {Weather Forecast}\ }\textbf {\bibinfo {volume} {{\bf 13}}},\ \bibinfo
  {pages} {1148} (\bibinfo {year} {1998})}\BibitemShut {NoStop}%
\bibitem [{\citenamefont {Levina}\ and\ \citenamefont
  {Montgomery}(2010)}]{montgo}%
  \BibitemOpen
  \bibfield  {author} {\bibinfo {author} {\bibfnamefont {G.}~\bibnamefont
  {Levina}}\ and\ \bibinfo {author} {\bibfnamefont {M.}~\bibnamefont
  {Montgomery}},\ }\href@noop {} {\bibfield  {journal} {\bibinfo  {journal}
  {Dokl. Earth Sci.}\ }\textbf {\bibinfo {volume} {{\bf 434}}},\ \bibinfo
  {pages} {1285} (\bibinfo {year} {2010})}\BibitemShut {NoStop}%
\bibitem [{\citenamefont {Markowski}\ \emph {et~al.}(1998)\citenamefont
  {Markowski}, \citenamefont {Straka}, \citenamefont {Rasmussen},\ and\
  \citenamefont {Blanchard}}]{marko}%
  \BibitemOpen
  \bibfield  {author} {\bibinfo {author} {\bibfnamefont {P.~M.}\ \bibnamefont
  {Markowski}}, \bibinfo {author} {\bibfnamefont {J.~M.}\ \bibnamefont
  {Straka}}, \bibinfo {author} {\bibfnamefont {E.~N.}\ \bibnamefont
  {Rasmussen}}, \ and\ \bibinfo {author} {\bibfnamefont {D.~O.}\ \bibnamefont
  {Blanchard}},\ }\href@noop {} {\bibfield  {journal} {\bibinfo  {journal}
  {Mont.\ Weath.\ Rev.}\ }\textbf {\bibinfo {volume} {{\bf 126}}},\ \bibinfo
  {pages} {2959} (\bibinfo {year} {1998})}\BibitemShut {NoStop}%
\bibitem [{\citenamefont {Xu}\ and\ \citenamefont {Wu}(2003)}]{xU_03}%
  \BibitemOpen
  \bibfield  {author} {\bibinfo {author} {\bibfnamefont {Y.}~\bibnamefont
  {Xu}}\ and\ \bibinfo {author} {\bibfnamefont {R.}~\bibnamefont {Wu}},\
  }\href@noop {} {\bibfield  {journal} {\bibinfo  {journal} {Adv. Atm. Phys.}\
  }\textbf {\bibinfo {volume} {{\bf 20}}},\ \bibinfo {pages} {940} (\bibinfo
  {year} {2003})}\BibitemShut {NoStop}%
\bibitem [{\citenamefont {Winn}\ \emph {et~al.}(1999)\citenamefont {Winn},
  \citenamefont {Hunyady},\ and\ \citenamefont {Aulich}}]{winn}%
  \BibitemOpen
  \bibfield  {author} {\bibinfo {author} {\bibfnamefont {W.}~\bibnamefont
  {Winn}}, \bibinfo {author} {\bibfnamefont {S.}~\bibnamefont {Hunyady}}, \
  and\ \bibinfo {author} {\bibfnamefont {G.}~\bibnamefont {Aulich}},\
  }\href@noop {} {\bibfield  {journal} {\bibinfo  {journal} {J. Geophys. Res.}\
  }\textbf {\bibinfo {volume} {{\bf 104(D18)}}},\ \bibinfo {pages} {22,067 }
  (\bibinfo {year} {1999})}\BibitemShut {NoStop}%
\bibitem [{\citenamefont {Moffatt}\ and\ \citenamefont
  {Tsinober}(1992)}]{tsinober}%
  \BibitemOpen
  \bibfield  {author} {\bibinfo {author} {\bibfnamefont {H.}~\bibnamefont
  {Moffatt}}\ and\ \bibinfo {author} {\bibfnamefont {A.}~\bibnamefont
  {Tsinober}},\ }\href@noop {} {\bibfield  {journal} {\bibinfo  {journal} {Ann.
  Rev. Fluid Mech.}\ }\textbf {\bibinfo {volume} {24}},\ \bibinfo {pages} {281}
  (\bibinfo {year} {1992})}\BibitemShut {NoStop}%
\bibitem [{\citenamefont {Mininni}\ \emph {et~al.}(2012)\citenamefont
  {Mininni}, \citenamefont {Rosenberg},\ and\ \citenamefont {Pouquet}}]{3072}%
  \BibitemOpen
  \bibfield  {author} {\bibinfo {author} {\bibfnamefont {P.}~\bibnamefont
  {Mininni}}, \bibinfo {author} {\bibfnamefont {D.}~\bibnamefont {Rosenberg}},
  \ and\ \bibinfo {author} {\bibfnamefont {A.}~\bibnamefont {Pouquet}},\
  }\href@noop {} {\bibfield  {journal} {\bibinfo  {journal} {J. Fluid Mech.}\
  }\textbf {\bibinfo {volume} {{\bf 699}}},\ \bibinfo {pages} {263 } (\bibinfo
  {year} {2012})}\BibitemShut {NoStop}%
\bibitem [{\citenamefont {Mininni}\ \emph {et~al.}(2011)\citenamefont
  {Mininni}, \citenamefont {Rosenberg}, \citenamefont {Reddy},\ and\
  \citenamefont {Pouquet}}]{hybrid2011}%
  \BibitemOpen
  \bibfield  {author} {\bibinfo {author} {\bibfnamefont {P.}~\bibnamefont
  {Mininni}}, \bibinfo {author} {\bibfnamefont {D.}~\bibnamefont {Rosenberg}},
  \bibinfo {author} {\bibfnamefont {R.}~\bibnamefont {Reddy}}, \ and\ \bibinfo
  {author} {\bibfnamefont {A.}~\bibnamefont {Pouquet}},\ }\href@noop {}
  {\bibfield  {journal} {\bibinfo  {journal} {Parallel Computing}\ }\textbf
  {\bibinfo {volume} {37}},\ \bibinfo {pages} {316} (\bibinfo {year}
  {2011})}\BibitemShut {NoStop}%
\bibitem [{\citenamefont {G\'omez}\ \emph {et~al.}(2005)\citenamefont
  {G\'omez}, \citenamefont {Mininni},\ and\ \citenamefont
  {Dmitruk}}]{gomez2005}%
  \BibitemOpen
  \bibfield  {author} {\bibinfo {author} {\bibfnamefont {D.~O.}\ \bibnamefont
  {G\'omez}}, \bibinfo {author} {\bibfnamefont {P.~D.}\ \bibnamefont
  {Mininni}}, \ and\ \bibinfo {author} {\bibfnamefont {P.}~\bibnamefont
  {Dmitruk}},\ }\href@noop {} {\bibfield  {journal} {\bibinfo  {journal}
  {Physica Scripta}\ }\textbf {\bibinfo {volume} {T116}},\ \bibinfo {pages}
  {123} (\bibinfo {year} {2005})}\BibitemShut {NoStop}%
\bibitem [{\citenamefont {Waite}\ and\ \citenamefont
  {Bartello}(2004)}]{waite2004}%
  \BibitemOpen
  \bibfield  {author} {\bibinfo {author} {\bibfnamefont {M.}~\bibnamefont
  {Waite}}\ and\ \bibinfo {author} {\bibfnamefont {P.}~\bibnamefont
  {Bartello}},\ }\href@noop {} {\bibfield  {journal} {\bibinfo  {journal} {J.
  Fluid Mech.}\ }\textbf {\bibinfo {volume} {517}},\ \bibinfo {pages} {281}
  (\bibinfo {year} {2004})}\BibitemShut {NoStop}%
\bibitem [{\citenamefont {Pouquet}\ and\ \citenamefont
  {Patterson}(1978)}]{patterson}%
  \BibitemOpen
  \bibfield  {author} {\bibinfo {author} {\bibfnamefont {A.}~\bibnamefont
  {Pouquet}}\ and\ \bibinfo {author} {\bibfnamefont {G.~S.}\ \bibnamefont
  {Patterson}},\ }\href@noop {} {\bibfield  {journal} {\bibinfo  {journal} {J.
  Fluid Mech.}\ }\textbf {\bibinfo {volume} {85}},\ \bibinfo {pages} {305}
  (\bibinfo {year} {1978})}\BibitemShut {NoStop}%
\bibitem [{\citenamefont {Sen}\ \emph {et~al.}(2012)\citenamefont {Sen},
  \citenamefont {Rosenberg}, \citenamefont {Pouquet},\ and\ \citenamefont
  {Mininni}}]{sen2012}%
  \BibitemOpen
  \bibfield  {author} {\bibinfo {author} {\bibfnamefont {A.}~\bibnamefont
  {Sen}}, \bibinfo {author} {\bibfnamefont {D.}~\bibnamefont {Rosenberg}},
  \bibinfo {author} {\bibfnamefont {A.}~\bibnamefont {Pouquet}}, \ and\
  \bibinfo {author} {\bibfnamefont {P.}~\bibnamefont {Mininni}},\ }\href@noop
  {} {\bibfield  {journal} {\bibinfo  {journal} {Phys. Rev. E}\ }\textbf
  {\bibinfo {volume} {86}},\ \bibinfo {pages} {036319} (\bibinfo {year}
  {2012})}\BibitemShut {NoStop}%
\bibitem [{\citenamefont {Hide}(2002)}]{hide_02}%
  \BibitemOpen
  \bibfield  {author} {\bibinfo {author} {\bibfnamefont {R.}~\bibnamefont
  {Hide}},\ }\href@noop {} {\bibfield  {journal} {\bibinfo  {journal} {Q. J.
  Roy. Met. Soc.}\ }\textbf {\bibinfo {volume} {{\bf 128}}},\ \bibinfo {pages}
  {1759 } (\bibinfo {year} {2002})}\BibitemShut {NoStop}%
\bibitem [{\citenamefont {Matthaeus}\ \emph {et~al.}(2008)\citenamefont
  {Matthaeus}, \citenamefont {Pouquet}, \citenamefont {Mininni}, \citenamefont
  {Dmitruk},\ and\ \citenamefont {Breech}}]{matthaeus}%
  \BibitemOpen
  \bibfield  {author} {\bibinfo {author} {\bibfnamefont {W.~H.}\ \bibnamefont
  {Matthaeus}}, \bibinfo {author} {\bibfnamefont {A.}~\bibnamefont {Pouquet}},
  \bibinfo {author} {\bibfnamefont {P.~D.}\ \bibnamefont {Mininni}}, \bibinfo
  {author} {\bibfnamefont {P.}~\bibnamefont {Dmitruk}}, \ and\ \bibinfo
  {author} {\bibfnamefont {B.}~\bibnamefont {Breech}},\ }\href@noop {}
  {\bibfield  {journal} {\bibinfo  {journal} {Phys. Rev. Lett.}\ }\textbf
  {\bibinfo {volume} {100}},\ \bibinfo {pages} {085003} (\bibinfo {year}
  {2008})}\BibitemShut {NoStop}%
\bibitem [{\citenamefont {Dombre}\ \emph {et~al.}(1986)\citenamefont {Dombre},
  \citenamefont {Frisch}, \citenamefont {Greene}, \citenamefont {H\'enon},
  \citenamefont {Mehr},\ and\ \citenamefont {Soward}}]{dombre}%
  \BibitemOpen
  \bibfield  {author} {\bibinfo {author} {\bibfnamefont {T.}~\bibnamefont
  {Dombre}}, \bibinfo {author} {\bibfnamefont {U.}~\bibnamefont {Frisch}},
  \bibinfo {author} {\bibfnamefont {J.~M.}\ \bibnamefont {Greene}}, \bibinfo
  {author} {\bibfnamefont {M.}~\bibnamefont {H\'enon}}, \bibinfo {author}
  {\bibfnamefont {A.}~\bibnamefont {Mehr}}, \ and\ \bibinfo {author}
  {\bibfnamefont {A.~M.}\ \bibnamefont {Soward}},\ }\href@noop {} {\bibfield
  {journal} {\bibinfo  {journal} {J. Fluid Mech.}\ }\textbf {\bibinfo {volume}
  {167}},\ \bibinfo {pages} {353} (\bibinfo {year} {1986})}\BibitemShut
  {NoStop}%
\bibitem [{\citenamefont {Billant}\ and\ \citenamefont
  {Chomaz}(2001)}]{chomaz}%
  \BibitemOpen
  \bibfield  {author} {\bibinfo {author} {\bibfnamefont {P.}~\bibnamefont
  {Billant}}\ and\ \bibinfo {author} {\bibfnamefont {J.-M.}\ \bibnamefont
  {Chomaz}},\ }\href@noop {} {\bibfield  {journal} {\bibinfo  {journal} {Phys.
  Fluids}\ }\textbf {\bibinfo {volume} {13}},\ \bibinfo {pages} {1645}
  (\bibinfo {year} {2001})}\BibitemShut {NoStop}%
\bibitem [{\citenamefont {Haynes}\ and\ \citenamefont
  {McIntyre}(1990)}]{MCI_90}%
  \BibitemOpen
  \bibfield  {author} {\bibinfo {author} {\bibfnamefont {P.}~\bibnamefont
  {Haynes}}\ and\ \bibinfo {author} {\bibfnamefont {M.}~\bibnamefont
  {McIntyre}},\ }\href@noop {} {\bibfield  {journal} {\bibinfo  {journal} {J.
  Atmosph. Sci.}\ }\textbf {\bibinfo {volume} {{\bf 42}}},\ \bibinfo {pages}
  {2021} (\bibinfo {year} {1990})}\BibitemShut {NoStop}%
\bibitem [{\citenamefont {Teitelbaum}\ and\ \citenamefont
  {Mininni}(2009)}]{teitelbaum}%
  \BibitemOpen
  \bibfield  {author} {\bibinfo {author} {\bibfnamefont {T.}~\bibnamefont
  {Teitelbaum}}\ and\ \bibinfo {author} {\bibfnamefont {P.}~\bibnamefont
  {Mininni}},\ }\href@noop {} {\bibfield  {journal} {\bibinfo  {journal} {Phys.
  Rev. Lett.}\ }\textbf {\bibinfo {volume} {103}},\ \bibinfo {pages} {014501}
  (\bibinfo {year} {2009})}\BibitemShut {NoStop}%
\bibitem [{\citenamefont {Galtier}\ \emph {et~al.}(1997)\citenamefont
  {Galtier}, \citenamefont {Politano},\ and\ \citenamefont
  {Pouquet}}]{galtier_mhd}%
  \BibitemOpen
  \bibfield  {author} {\bibinfo {author} {\bibfnamefont {S.}~\bibnamefont
  {Galtier}}, \bibinfo {author} {\bibfnamefont {H.}~\bibnamefont {Politano}}, \
  and\ \bibinfo {author} {\bibfnamefont {A.}~\bibnamefont {Pouquet}},\
  }\href@noop {} {\bibfield  {journal} {\bibinfo  {journal} {Phys. Rev. Lett.}\
  }\textbf {\bibinfo {volume} {{\bf 79}}},\ \bibinfo {pages} {2807} (\bibinfo
  {year} {1997})}\BibitemShut {NoStop}%
\bibitem [{\citenamefont {Teitelbaum}\ and\ \citenamefont
  {Mininni}(2011)}]{pablo_thomas}%
  \BibitemOpen
  \bibfield  {author} {\bibinfo {author} {\bibfnamefont {T.}~\bibnamefont
  {Teitelbaum}}\ and\ \bibinfo {author} {\bibfnamefont {P.}~\bibnamefont
  {Mininni}},\ }\href@noop {} {\bibfield  {journal} {\bibinfo  {journal} {Phys.
  Fluids}\ }\textbf {\bibinfo {volume} {{\bf 23}}},\ \bibinfo {pages} {065105}
  (\bibinfo {year} {2011})}\BibitemShut {NoStop}%
\bibitem [{\citenamefont {Mininni}\ and\ \citenamefont
  {Pouquet}(2010{\natexlab{a}})}]{1536a}%
  \BibitemOpen
  \bibfield  {author} {\bibinfo {author} {\bibfnamefont {P.}~\bibnamefont
  {Mininni}}\ and\ \bibinfo {author} {\bibfnamefont {A.}~\bibnamefont
  {Pouquet}},\ }\href@noop {} {\bibfield  {journal} {\bibinfo  {journal} {Phys.
  Fluids}\ }\textbf {\bibinfo {volume} {22}},\ \bibinfo {pages} {035105}
  (\bibinfo {year} {2010}{\natexlab{a}})}\BibitemShut {NoStop}%
\bibitem [{\citenamefont {Mininni}\ and\ \citenamefont
  {Pouquet}(2010{\natexlab{b}})}]{1536b}%
  \BibitemOpen
  \bibfield  {author} {\bibinfo {author} {\bibfnamefont {P.}~\bibnamefont
  {Mininni}}\ and\ \bibinfo {author} {\bibfnamefont {A.}~\bibnamefont
  {Pouquet}},\ }\href@noop {} {\bibfield  {journal} {\bibinfo  {journal} {Phys.
  Fluids}\ }\textbf {\bibinfo {volume} {22}},\ \bibinfo {pages} {035106}
  (\bibinfo {year} {2010}{\natexlab{b}})}\BibitemShut {NoStop}%
\bibitem [{\citenamefont {Pouquet}\ and\ \citenamefont
  {Mininni}(2010)}]{trieste}%
  \BibitemOpen
  \bibfield  {author} {\bibinfo {author} {\bibfnamefont {A.}~\bibnamefont
  {Pouquet}}\ and\ \bibinfo {author} {\bibfnamefont {P.}~\bibnamefont
  {Mininni}},\ }\href@noop {} {\bibfield  {journal} {\bibinfo  {journal} {Phil.
  Trans. Roy. Soc.}\ }\textbf {\bibinfo {volume} {368}},\ \bibinfo {pages}
  {1635} (\bibinfo {year} {2010})}\BibitemShut {NoStop}%
\bibitem [{\citenamefont {Kimura}\ and\ \citenamefont
  {Herring}(2012)}]{kimura12}%
  \BibitemOpen
  \bibfield  {author} {\bibinfo {author} {\bibfnamefont {Y.}~\bibnamefont
  {Kimura}}\ and\ \bibinfo {author} {\bibfnamefont {J.~R.}\ \bibnamefont
  {Herring}},\ }\href@noop {} {\bibfield  {journal} {\bibinfo  {journal} {J.
  Fluid Mech.}\ }\textbf {\bibinfo {volume} {698}},\ \bibinfo {pages} {19}
  (\bibinfo {year} {2012})}\BibitemShut {NoStop}%
\bibitem [{\citenamefont {Godeferd}\ and\ \citenamefont
  {Staquet}(2003{\natexlab{b}})}]{godeferd}%
  \BibitemOpen
  \bibfield  {author} {\bibinfo {author} {\bibfnamefont {F.}~\bibnamefont
  {Godeferd}}\ and\ \bibinfo {author} {\bibfnamefont {C.}~\bibnamefont
  {Staquet}},\ }\href@noop {} {\bibfield  {journal} {\bibinfo  {journal} {J.
  Fluid Mech.}\ }\textbf {\bibinfo {volume} {486}},\ \bibinfo {pages} {115 }
  (\bibinfo {year} {2003}{\natexlab{b}})}\BibitemShut {NoStop}%
\bibitem [{\citenamefont {Waite}(2011)}]{waite2011}%
  \BibitemOpen
  \bibfield  {author} {\bibinfo {author} {\bibfnamefont {M.~L.}\ \bibnamefont
  {Waite}},\ }\href@noop {} {\bibfield  {journal} {\bibinfo  {journal} {Phys.
  of Fluids}\ }\textbf {\bibinfo {volume} {23}},\ \bibinfo {pages} {066602}
  (\bibinfo {year} {2011})}\BibitemShut {NoStop}%
\bibitem [{\citenamefont {Bartello}\ and\ \citenamefont
  {Tobias}(2012)}]{bartello_13}%
  \BibitemOpen
  \bibfield  {author} {\bibinfo {author} {\bibfnamefont {P.}~\bibnamefont
  {Bartello}}\ and\ \bibinfo {author} {\bibfnamefont {S.~M.}\ \bibnamefont
  {Tobias}},\ }\href@noop {} {\bibfield  {journal} {\bibinfo  {journal}
  {submitted to J. Fluid Mech.}\ } (\bibinfo {year} {2012})}\BibitemShut
  {NoStop}%
\bibitem [{\citenamefont {Waite}(2012)}]{waite_2012}%
  \BibitemOpen
  \bibfield  {author} {\bibinfo {author} {\bibfnamefont {M.~L.}\ \bibnamefont
  {Waite}},\ }\href@noop {} {Direct numerical
  simulations of laboratory-scale stratified turbulence. Submitted to Modeling Atmospheric and Oceanic Flows: Insights from
  Laboratory Experiments and Numerical Simulations. Geophysical Monograph,
  AGU.}\ (\bibinfo  {publisher} {T. von Larcher and P. Williams (eds.)},\
  \bibinfo {year} {2012})\BibitemShut {NoStop}%
\end{thebibliography}%

\end{document}